\documentclass[pra, twocolumn, showpacs]{revtex4}
\usepackage{latexsym}
\usepackage{amssymb}
\usepackage{amsmath}
\usepackage{exscale}
\usepackage{bm, graphics}
\usepackage{psfrag}
\usepackage{graphicx}
\providecommand{\abs}[1]{|#1|}
\newcommand{\Nabla}{\mbox{\boldmath $\nabla$}}
\newcommand{\D}{\mbox{\rm d}}
\newcommand{\mbb}[1]{\mbox{\boldmath $#1$}} 
\begin{document}
\author{Mikayel  Khanbekyan, Ludwig Kn\"oll, and Dirk-Gunnar Welsch}
\affiliation{Theoretisch-Physikalisches Institut,
Friedrich-Schiller-Universit\"{a}t Jena, Max-Wien-Platz 1, D-07743,
Jena, Germany}
\title{
%%Three-dimensional quantum-optical input-output
%%relations for a multilayer dielectric plate
%The quantized electromagnetic field in the presence of
%dispersing and absorbing planar multilayers
Input-output relations at
dispersing and absorbing planar multilayers 
for the quantized electromagnetic field containing
%propagating and
evanescent components  
}
\date{\today}
\begin{abstract} 
By using the Green-function concept of quantization of the
electromagnetic field in dispersing and absorbing media,
the quantized field in the presence of
a dispersing and absorbing  dielectric multilayer plate
is studied. Three-dimensional input-output
relations are derived for both amplitude operators
in the ${\bf k}$-space and the field operators 
in the coordinate space. The conditions are discussed,
under which the input-output relations can be expressed
in terms of bosonic operators. The theory applies to
both (effectively) free fields and fields,
created by active atomic sources inside and/or outside
the plate, including also evanescent-field components.
\end{abstract}
\maketitle
\reversemarginpar

\section{Introduction}
\label{sec1}

Research on quantum statistical properties of light has been
a subject of intense investigations and discussions.
In view of the very wide-spread potential applications,
the influence of material bodies on the quantum features of light
must be thoroughly investigated.
Typical quantum effects that are closely related to the change
of the quantum vacuum due to the presence of material bodies
are the Casimir effect \cite{Casimir} (for a review, see
\cite{Bordag}) and the
Purcell effect \cite{Purcell} (see also \cite{Milonni}). 
Though there has been a large body of work on the problem
of quantization of the electromagnetic field
in dielectric media (for a review, see \cite{Luks02}),
most work has been concentrated on nonabsorbing media. 
Roughly speaking, there have been two routes to
treat radiation in absorbing media. In the first,
attention has been restricted to equilibrium field correlation
functions that are calculated by employing the
dissipation-fluctuation theorem \cite{Agarwal11}.
In the second, explicit field quantization has been performed
(for a review, see \cite{Knoell01}).

A consistent quantization scheme for the electromagnetic field
in bulk material has been given by Huttner and Barnett
\cite{Huttner46}. Using the mesoscopic Hopfield model \cite{Hopfield}
for a dielectric, they have diagonalized
the bilinear Hamiltonian of the system
that consists of the electromagnetic field, a
harmonic-oscillator polarization field, and
an infinite set of harmonic-oscillator reservoir
fields responsible for absorption.
By starting directly with the phenomenological Maxwell
equations for the macroscopic electromagnetic field,
the scheme can naturally be extended
to arbitrary inhomogeneous media \cite{Gruner53} 
characterized by a space- and frequency-dependent
complex permittivity that satisfies the Kramers-Kronig
relations. The theory
is based on a source-quantity representation
of the electromagnetic field, where the field operators are expressed 
in terms of a continuous set of fundamental Bose fields
via the Green tensor of the classical 
problem (for details, see also \cite{Knoell01}).

The Green-tensor formalism
is well suited to study
the behavior of the quantized electromagnetic field in the presence
of dispersing and absorbing bodies.
In particular, it has successfully been applied to
the study of input-output relations \cite{Gruner54,Scheel3D},
the spontaneous emission
\cite{Ho01,Bondarev02,Ho03},
the resonant energy exchange \cite{Ho02}, the
resonant dipole-dipole interaction \cite{Ho02a},
and the Casimir effect \cite{Tomas02}, and also various geometries
(including multilayer structures) have been considered.
It is worth noting that the theory not only takes into
account material absorption in a consistent way but it
also includes automatically evanescent-field
contributions generated, e.g., by radiating atoms that
are very close to the bodies under consideration.

For the particular case of a dispersing and absorbing
dielectric structure in vacuum, without active light sources
at any finite distance from the structure
(and without active light sources inside the structure),
a ratherinvolved hybrid quantization scheme
has been proposed \cite{DiStefano01} and used to
study the problem of three-dimensional input-output
relations \cite{Savasta65, Savasta19}.  
In the scheme, the whole space is divided into
two regions, namely a dielectric scattering region
and a vacuum region that surrounds the scattering region.
The quantization in the vacuum region is performed
on the basis of a mode decomposition 
of the electromagnetic-field operators
in a similar way as in vacuum quantum electrodynamics,
whereas the contribution of the dielectric scattering region
is taken into account by applying the Green tensor
formalism \cite{Gruner53}. 
The argument is \cite{Savasta19} that the
Green tensor formalism alone is not complete, because it
does not explicitly treat the boundary of the
medium with free space and thus does not describe
explicitly the scattering and emission processes
which are usually experimentally investigated.
This is of course not the case as the above mentioned applications
\cite{Gruner54,Scheel3D,Ho01,Bondarev02,Ho03,Ho02,Tomas02}
and the following study of the three-dimensional input-output
relations at multilayer plates clearly show.
Moreover, when there are active light sources at {\em finite
distances} from the scattering region, then the incoming field
contains both propagating and evanescent components.
The latter ones, which are typically observed for small distances,
are not included in the hybrid quantization scheme \cite{DiStefano01}
and the input-output relations derived from it \cite{Savasta19}.
However, they are automatically included in the
Green tensor formalism \cite{Gruner53}.
 
In the following we consider the problem of the
three-di\-mensional input-output relations for
the el\-ec\-t\-ro\-magnetic-field operators at
the boundaries of a dielectric planar multilayer structure
in more detail, by applying the Green tensor formalism and
extending our previous one-dimensional analysis \cite{Gruner54}
to three dimensions. 
To be quite general, we allow (i) for 
the presence of active light sources
at arbitrary positions
inside and/or outside the multilayer plate
and thus (ii) for both propagating-field
components and evanescent-field components.
In particular, in cavity QED the active sources are
typically inside the plate, and 
the outgoing fields are not only determined
with the incoming fields and the noise fields (unavoidably associated
with material absorption) but also with the fields generated
by the sources inside the plate. On the contrary, in
optical near-field microscopy the active (probe) sources
are typically outside the plate, but near its surface.

Employing the well-known
three-dimensional Green tensor
for a multilayer plate \cite{Tomas95}, we introduce
amplitude operators for the input and output fields
%and 
as well as for
the fields inside the plate
and derive input-output relations both in the two-dimensional Fourier
space and in the coordinate space. 
Finally, the problem of introduction of 
bosonic input and output operators
in the two-dimensional Fourier space
is considered in detail.

The paper is organized as follows.
In Sec.~\ref{sec:ba} the quantization scheme is briefly
summarized and the Green tensor for a multilayer
plate is introduced. In Sec.~\ref{sec:gr} 
the generally valid input-output relations are derived,
and the problem of formulating input-output relations
for bosonic field operators is studied. A summary and some
concluding remarks are given in Sec.~\ref{sec:co}
followed by an appendix, in which relevant commutation
relations are derived. 

%%%%%%%%%%%%%%%%%%%%%%%%%%%%%%%%%%%%%%%%%%%%%%%%%%%%%%%%%%%%%%%%%%%%%%%%%
%%%%%%%%%%%%%%%%%%%%%%%%%%%%%%%%%%%%%%%%%%%%%%%%%%%%%%%%%%%%%%%%%%%%%%%%%

\section{\label{sec:ba}Basic equations}

To study optical fields interacting
with active sources in the presence of dispersing and
absorbing (linear) dielectric bodies, we first note that on 
a length scale that is large compared with interatomic
distances in the bodies, the effect of the bodies
can be described within the frame of macroscopic
Maxwell equations in terms of a spatially varying
permittivity which is a complex function of frequency.
This concept, which is widely used in classical optics
also applies in quantum optics.

%%%%%%%%%%%%%%%%%%%%%%%%%%%%%%%%%%%%%%%%%%%%%%%%%%%%%%%%%%%%%%%%%%%%%%%

\subsection{Field quantization in media}
\label{sec:ba.1}

Let us assume that the active light sources are neutral atoms
and consider an arbitrarily inhomogeneous medium characterized
by a permittivity
\begin{equation}
      \label{1}
      \varepsilon({\bf r},\omega) = \varepsilon'({\bf r},\omega)
      + i\varepsilon''({\bf r},\omega),
      \end{equation}
where the real part $\varepsilon'({\bf r},\omega)$ and the imaginary part
$\varepsilon''({\bf r},\omega)$ are necessarily related to each other
through the Kramers-Kronig relations, due to the causality principle.
The motion of the atoms and the medium-assisted electromagnetic field
is then governed by the multipolar-coupling Hamiltonian \cite{Knoell01}
\begin{widetext}
\begin{eqnarray}
      \label{2}
      \hat{H} &=& \int\! \D^3{r} \int_0^\infty\! \D\omega
      \,\hbar\omega\,\hat{\bf f}^{\dagger}({\bf r},\omega)\cdot
      \hat{\bf f}({\bf r},\omega)
      + \!\sum_{A,\alpha_A}\! \frac{1}{2m_{\alpha_A}}
      \bigg\{ \hat{{\bf p}}_{\alpha_A} +
      q_{\alpha_A}\int_0^1\!{\rm d}\lambda\,\lambda
      \left( \hat{\bf r}_{\alpha_A}\!-\!
      {\bf r}_A \right) \times \hat{\bf B}\left[ {\bf r}_A\!+\!
      \lambda \left( \hat{\bf r}_{\alpha_A} \!-\!{\bf r}_A \right) \right]
      \!\!\bigg\}^2
\nonumber \\[1ex]&&
      +\,\frac{1}{2\epsilon_0}
      \sum_{A,A'} \int \D^3{r} \,
      \hat{\bf P}_{A}({\bf r})\cdot \hat{\bf P}_{A'}({\bf r})
      -\sum_A \int \D^3{r}\, \hat{{\bf P}}_{A}({\bf r})\cdot
      \hat{\bf E}({\bf r}),
      \end{eqnarray}
\end{widetext}
where $A$ numbers the atoms, and $\alpha_A$ numbers the charged
particles inside the $A$-th atom. Further, $\hat{\bf r}_{\alpha_A}$ and
$\hat{\bf p}_{\alpha_A}$ are respectively the operators
of coordinates and canonical momenta of the particles, and
\begin{equation}
      \label{3}
      \hat{{\bf P}}_{A}({\bf r}) =
      \! \sum_{\alpha_A} q_{\alpha_A}
      \left(\hat{{\bf r}}_{\alpha_A}\!-\! {\bf r}_{A}\right)
      \int_0^1  \D\lambda
      \delta \left[{\bf r}\!-\!{\bf r}_{A}\!
      -\!\lambda\left(\hat{{\bf r}}_{\alpha_A}\!
      -\!{\bf r}_{A}\right)\right]
      \end{equation}
is the operator of the polarization of the $A$-th atom at position
${\bf r}_A$. The $\hat{\bf f}({\bf r},\omega)$
[and $\hat{\bf f}^\dagger({\bf r},\omega)$]
are bosonic field operators that play the role of
fundamental variables of the electromagnetic
field and the medium, including a reservoir necessarily
associated with material absorption.
The commutation relations for the operators
${\bf \hat {f}} ({\bf {r }}, \omega )$  and
${\bf \hat {f}}^\dagger ({\bf {r }}, \omega )$ are
\begin{equation}
      \label{4}
      \bigl[\hat{f}_{\mu} ({\bf r}, \omega),
      \hat{f}_{\mu'} ^{\dagger } ({\bf r }',  \omega ') \bigr]
      = \delta _{\mu \mu'}\delta (\omega - \omega  ')
      \delta ^{(3)}({\bf r} - {\bf r }'),
\end{equation}
\begin{equation}
      \label{5}
      \bigl[\hat {f}_{\mu} ({\bf r}, \omega),
      \hat {f}_{\mu'}  ({\bf r }', \omega ') \bigr] = 0,
\end{equation}      
where the Greek letters label the Cartesian coordinates $x, \,y, \,z$.
In Eq.~(\ref{2}), 
the operators $\hat{\bf B}({\bf r})$
and $\hat{\bf E}({\bf r})$ of the medium-assisted electromagnetic
field are expressed in terms of the fundamental variables as
follows:
\begin{equation}
      \label{6}
      \hat{\bf B}({\bf r}) = \int_0^\infty \D\omega\,
      \hat{\bf B}({\bf r},\omega) + \mbox{H.c.},
      \end{equation}
      \begin{equation}
      \label{7}
      \hat{\bf E}({\bf r}) = \int_0^\infty \D\omega\,
      \hat{\bf E}({\bf r},\omega) + \mbox{H.c.},
      \end{equation}
\begin{equation}
      \label{8}
      \hat{\bf B}({\bf r},\omega)
      = (i\omega)^{-1}\Nabla\times\hat{\bf E}({\bf r},\omega),
      \end{equation}
\begin{eqnarray}
      \label{9}
\lefteqn{
      \hat{\bf E}({\bf r},\omega)
      = i \mu _0 \sqrt{\frac {\hbar \epsilon _0}{\pi}}\,\omega^2
}
\nonumber\\&&\times\;
      \int \D^3r'\sqrt{\varepsilon''({\bf r}',\omega)}\,
      \mbb{G}({\bf r},{\bf r}',\omega)\cdot\hat{\bf f}({\bf r}',\omega),
\end{eqnarray}
where the integration should be performed over all space, and
$ \mbb{G}({\bf r},{\bf r}',\omega)$ is the classical Green tensor,
which can be found from the equation
\begin{equation}
      \label{10}
      \Nabla \times \Nabla  \times  \mbb{G}  ({\bf r }, {\bf r }', \omega)
      - \frac {\omega ^2 } {c^2} \,\varepsilon ( {\bf r } ,\omega)
      \mbb{G}  ({\bf r}, {\bf r }', \omega)
      =  {\bf \delta} ^{(3)}  ({\bf r }-{\bf r }')
      \end{equation}
together with appropriate boundary conditions at infinity.
Equations (\ref{6} - \ref{9}) can be considered as generalization
of the familiar mode decomposition that would apply if dispersion
and absorption could be disregarded.
Instead of dealing with equations of motion for mode operators,
equations of motion
for the fields $\hat{\bf f}({\bf r},\omega)$ must be treated.

It should be pointed out
that since the real part $\varepsilon'({\bf r},\omega)$
and the imaginary part $\varepsilon''({\bf r},\omega)$
of the permittivity
are related to each other through the Kramers-Kronig relations,
$\varepsilon''({\bf r},\omega)$ cannot vanish identically
for really existing media. Clearly, $\varepsilon''({\bf r'},\omega)$
can be very small, so that $\sqrt{\varepsilon''({\bf r}',\omega)}\,
\mbb{G}({\bf r},{\bf r}',\omega)$ in Eq.~(\ref{9}) is very
small in certain areas of space (${\bf r}'$).
However, this statement says nothing about the magnitude of the
total integral over the coordinate ${\bf r}'$, because of the
following integral relation for the classical Green tensor
\begin{eqnarray}
      \label{11}
      \int \D^3{r}'\frac{\omega^2}{c^2}\varepsilon''({\bf r}',\omega)
      G_{\mu\mu'}({\bf r},{\bf r}',\omega)
      G_{\mu''\mu'}^*({\bf r''},{\bf r'},\omega)
\nonumber \\ 
      = {\rm Im}\,\left[G_{\mu\mu''}({\bf r},{\bf r}'',\omega)\right],
      \end{eqnarray}
where we have adopted the convention of summation over repeated
vector-component indices. In particular, this relation enables one
to include also vacuum-like areas in the consideration.
Thus, all the calculations are to be performed by assuming a permittivity
close to unity with a small but finite imaginary part in those areas,
and at the end the permittivity may be set equal to unity.
In practice, experimental realization of (macroscopic) vacuum areas is
of course fictional.

%%%%%%%%%%%%%%%%%%%%%%%%%%%%%%%%%%%%%%%%%%%%%%%%%%%%%%%%%%%%%%%%%%%%%%%%

\begin{figure}[tbh]
%	\unitlength=1in
%	\psfragscanon
%	\psfrag{a0_in} [l] {$\hat{E}_{in}^{(0)}$}
%	\psfrag{an_in} [l] {$\hat{E}_{in}^{(n)}$}
%	\psfrag{a0_out} [l] {$\hat{E}_{out}^{(0)}$}
%	\psfrag{an_out} [l] {$\hat{E}_{out}^{(n)}$}
%	\psfrag{e}[l]{$\varepsilon _0$}
%	\psfrag{ei}[l]{$\varepsilon _j$}
%	\psfrag{en}[l]{$\varepsilon _n$}
%	\psfrag{1}[c][r]{$z=0^-$}
%	\psfrag{2}[c][r]{$z=0^+$}
%	\psfrag{di}[l]{$d_j$}
%	\psfrag{z}[l]{$z$}
%	\psfrag{3}[r]{$\cdots$}
%	\psfrag{4}[l]{$\cdots$}
%\includegraphics[width=3.8in]{pic-t.eps}
\centerline{
\includegraphics[width=3.4in]{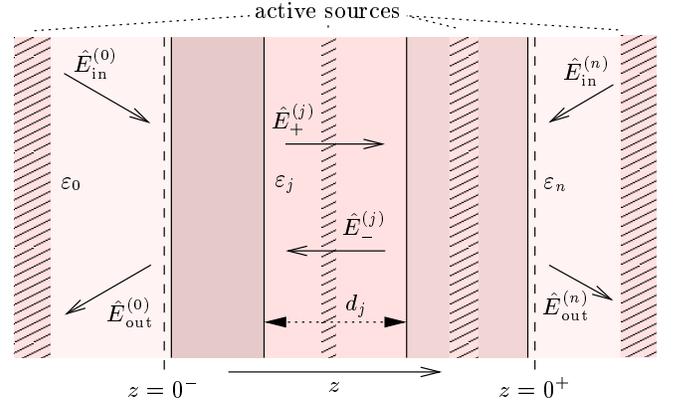}}
\caption{\label{fig}Scheme of the multilayer dielectric plate.
The hatched regions indicate the presence of active light sources. 
%The dashed lines represent the light sources inside and outside 
%the multilayer structure.
}
\end{figure}

%%%%%%%%%%%%%%%%%%%%%%%%%%%%%%%%%%%%%%%%%%%%%%%%%%%%%%%%%%%%%%%%%%%%%%%%%

\subsection{Multilayered planar structures}
\label{sec:ba.2}

As mentioned, the quantization scheme
is valid for an arbitrary space dependence of the permittivity.
Here, we consider a multilayered planar structure
(Fig.~\ref{fig}), whose permittivity is defined in a stepwise fashion
(the $z$-direction is chosen to be perpendicular to the layers):
\begin{equation}
      \label{12}
      \varepsilon (\omega, z)
      = \sum _{j=0} ^n \lambda _j (z)\varepsilon _j (\omega),
      \end{equation}
where
\begin{equation}
      \label{13}
      \lambda _j (z) =
        \left\{ \begin{array}{cl}
                1, \, & \mbox{if $z \in j$th layer},\\[1ex]
                0, \, & \mbox{otherwise},
        \end{array} \right.
      \end{equation}
and $\varepsilon _j (\omega)$ is the (complex) permittivity
of the $j$-th layer. In the above, the index $j$ labels the
region on the left of the plate \mbox{($j$ $\!=$ $\!0$)},
the region on the right of the plate \mbox{($j$ $\!=$ $\!n$)},
and the layers of the plate \mbox{($j$ $\!=$ $\!1,\dots ,n-1$)}.
For simplicity, we express the $z$-coordinate dependence
in shifted coordinate systems, introduced in each layer separately,
so that the range of the $z$-coordinate is taken to be
\mbox{$-\infty$ $\!<$ $\!z$ $\!<0$}
for the region  on the left of the plate \mbox{($j$ $\!=$ $\!0$)},
\mbox{$0$ $\!<$ $\!z$ $\!<\infty$}
for the region  on the right of the plate \mbox{($j$ $\!=$ $\!n$)},
and \mbox{$0$ $\!<$ $\!z$ $\!<$ $\!d_j$} for the $j$-th layer
of the plate with thickness $d_j$
\mbox{($j$ $\!=$ $\!1,\dots , n-1$)}.

Exploiting the translational symmetry in the \mbox{$(xy)$-p}la\-ne,
we may represent the Green tensor as a two-di\-men\-sional
Fourier integral
\begin{equation}
      \label{14}
      \mbb{G}^{(jj')}({\bf r},{\bf r'},\omega )
      =  \frac {1} {(2 \pi)^2} \!\int\! \D^2{k} \,
      e^{i{\bf k} ({\bm  \rho}-{\bm \rho'})}\,
      \mbb{G}^{(jj')}(z, z', {\bf k}, \omega ),
      \end{equation}
where \mbox{$\bm \rho$ $\!=$ $\!(x,y)$},
and \mbox{${\bf k }$ $\!=$ $\!(k_x, k_y)$} is the wave
vector parallel to the layers. The notations
${\bf G}^{(jj')}({\bf r}, {\bf r}', \omega )$
and ${\bf G}^{(jj')}(z, z', {\bf k}, \omega )$ indicate
that $z$ varies in the $j$-th layer and $z'$ in the
$j'$-th layer. Inserting Eq.~(\ref{14})
into Eq.~(\ref{9}), we may write the electric-field
operator ${\hat{\bf E}} ^{(j)}( {\bf r} , \omega )$ as a
two-fold Fourier transform,
\begin{equation}
      \label{15}
      {\hat{\bf E}} ^{(j)}( {\bf r} , \omega )
      = \frac {1} {(2 \pi )^2} \int \D^2{k}\,
      e^{i{\bf k } {\bm \rho} }
      \hat{\bf E}  ^{(j)}(z, {\bf k  }, \omega),
      \end{equation}
where
\begin{eqnarray}
      \label{16}
      \lefteqn{\hat{\bf E}^{(j)}(z, {\bf k  }, \omega) = i \omega \mu _0\nonumber} \\[1ex]
      && \times \sum _{j'=0}^n \int _{[j']} \D z'\,
      \mbb{G}^{(jj')}(z, z', {\bf k  }, \omega)\cdot
      \hat {\bf j}^{(j')} (z', {\bf k  }, \omega )
      \end{eqnarray}
($[j']$ indicates integration over the $j'$-th region).
Here, the Fourier transformed current operators  
\begin{equation}
      \label{17}
      \hat {\bf j}^{(j)} (z, {\bf k  }, \omega ) =
      \omega \sqrt { \frac {\hbar \epsilon _0} {\pi }\,
      \varepsilon '' _j (\omega) } \,
      \hat {\bf f}^{(j)} (z, {\bf k  }, \omega )
      \end{equation}
have been introduced, where the 
operators $\hat {\bf f}^{(j)} (z, {\bf k  }, \omega )$
are the Fourier transforms of the original bosonic field operators
$\hat {\bf f}^{(j)} ({\bf r}, \omega ) $,
\begin{equation}
      \label{18}
      \hat {\bf f}^{(j)} ({\bf r}, \omega )
      = \frac {1} {(2 \pi )^2} \int \D^2{k}\,
      e^{i{\bf k } {\bm \rho} }\,
      \hat {\bf f}^{(j)} (z, {\bf k  }, \omega ).
      \end{equation}

The Green tensor $\mbb{G}^{(jj')} (z, z', {\bf k  }, \omega)$
taken from Ref.~\cite{Tomas95} may be written as
\begin{eqnarray}
      \label{19}
      \lefteqn{\mbb{G}^{(jj')}(z, z', {\bf k  },  \omega )}
\nonumber\\[1ex]&&
      = - \,{\bf e}_z \,\frac {\delta_{jj'}} {k_j ^2} \,{\bf e} _z\,
      \delta (z-z') +   \mbb{g}^{(jj')}(z, z', {\bf k  },  \omega ),
\end{eqnarray}
where
\begin{eqnarray}
      \label{20}
\lefteqn{      
      \mbb{g}^{(jj')}(z, z', {\bf k  },  \omega )
}
\nonumber\\[1ex]&&
      =
      \frac {i} {2} \sum _{q=p,s}\sigma _q \left[
      \bm {\mathcal{E}} ^{j>} _{q}  (z, {\bf k  },  \omega )\,
      \Xi^{jj'} _{q}  \bm{\mathcal{E}} ^{j'<} _{q}  (z', -{\bf k  },  \omega)
      \Theta (j\!-\!j')
      \right.
\nonumber \\[1ex] && 
      \left.
      \,+\,
      \bm{\mathcal{E}}^{j<} _{q}  (z, {\bf k  },  \omega )\,
      \Xi^{j'j} _{q} \bm{\mathcal{E}} ^{j'>} _{q} (z', -{\bf k  },  \omega )
      \Theta (j'-j)
      \right] 
      \end{eqnarray}
(\mbox{$\sigma _p$ $\!=$ $\!1$}, \mbox{$\sigma _s$ $\!=$ $\!-1$}).
Note that for \mbox{$j$ $\!=$ $\!j'$} one should write
\mbox{$\Theta (z$ $\!-$ $\!z')$}
instead of \mbox{$\Theta (j$ $\!-$ $\!j')$}
[and \mbox{$\Theta (z'$ $\!-$ $\!z)$} instead of
\mbox{$\Theta (j'$ $-$ $\!j)$}].
In Eq.~(\ref{20}), the functions
$\bm{{\mathcal{E}}} ^{j>} _{q}  ({\bf k  },  \omega , z)$
and $ \bm{{\mathcal{E}}} ^{j<} _{q}  ({\bf k  },  \omega , z)$
denote  waves of unit strength traveling, respectively,
rightward and leftward in the $j$-th layer and being reflected
at the boundary,
\begin{eqnarray}
      \label{21}
  \lefteqn{    
\bm{\mathcal{E}} ^{(j)>} _{q}  (z, {\bf k  },  \omega )
}\nonumber\\ &&
\, = \,
      {\bf e}^{(j)}_{q+} ({\bf k  })  e^{i \beta _j (z-d _j)}
      + r^q _{j/n} {\bf e}^{(j)}_{q-} ({\bf k  })
      e^{-i \beta _j (z-d _j)}, \quad
      \end{eqnarray}
\begin{eqnarray}
      \label{22}
      \bm{{\mathcal{E}}} ^{(j)<} _{q}  (z, {\bf k  },  \omega )\, =\,
      {\bf e}^{(j)}_{q-} ({\bf k  })  e^{-i \beta _j z}
      + r^q _{j/0} {\bf e}^{(j)}_{q+} ({\bf k  })  e^{i \beta _j z},
\end{eqnarray}
and
\begin{equation}
      \label{23}
      \Xi^{jj'} _{q}
      \,=\,
      \frac{1}{\beta_n t^q_{0/n}}\,
      \frac{t^q_{0/j}e^{ i \beta _j d _j}}{D _{q j}}\,
      \frac{t^q_{n/j'}e^{ i \beta _{j'} d _{j'}}}{D _{q j'}}\,,
      \end{equation}
where
\begin{eqnarray}
      \label{24}
      D_{qj} &=& 1 - r^q _{j/0} r^q _{j/n} e^{2 i \beta _j d _j}
      \end{eqnarray}
\mbox{($d_0$ $\!=$ $\!d_n$ $\!=$ $\!0$)}. Here,
\begin{equation}
      \label{25}
      \beta _j = \sqrt { k_j ^2 - k^2}
      = \beta _j ' + i \beta _j ''
      \qquad  ( \beta_j ', \beta_j '' \geq 0 )
      \end{equation}
($k$ $\!=$ $|{\bf k}|$), where
\begin{equation}
      \label{26}
      k _j = \sqrt {\varepsilon _j (\omega)} \,\frac {\omega} {c}
      = k _j ' + i k _j ''
      \qquad  ( k _j ', k _j '' \geq 0 ),
      \end{equation}
and \mbox{$t_{j/j'}$ $\!=$ $\!(\beta_j/\beta_{j'})t_{j'/j}$}
and $r_{j/j'}$ are, respectively, the transmission and reflection
coefficients between the layers $j'$ and $j$.
Finally, the unit vectors ${\bf e}^{(j)} _{q\pm} ({\bf k  })$
in Eqs.~(\ref{21}) and (\ref{22})
are the polarization unit vectors for TE \mbox{($q$ $\!=$ $\!s$)}
and TM \mbox{($q$ $\!=$ $\!p$)} waves,
\begin{eqnarray}
      \label{27}
      {\bf e}^{(j)}_{s\pm}({\bf k})
      &=&
      \frac{{\bf k}}{k} \times  {\bf e} _z \,,
      \\[1ex]
      \label{28}
      {\bf e}^{(j)}_{p\pm}({\bf k})
      &=&
      \frac {1} {k_j} \left(\mp \beta _j\frac{{\bf k}}{k}
      + k {\bf e} _z \right).
      \end{eqnarray}
The `propagation constant' $\beta _j$ determines the propagation
behavior in $z$-direction of the waves in the $j$-th region.
Note that in case of vacuum the waves
are propagating only for \mbox{$\beta _j$ $\!=$ $\!\beta _j'$}
(i.e., \mbox{$\omega/c$ $\!>$ $\!k$}). They are evanescent
for \mbox{$\beta _j$ $\!=$ $\!i\beta _j''$}
(i.e., \mbox{$\omega/c$ $\!\leq$ $\!k$}).

%%%%%%%%%%%%%%%%%%%%%%%%%%%%%%%%%%%%%%%%%%%%%%%%%%%%%%%%%%%%%%%%%%%%%%
%%%%%%%%%%%%%%%%%%%%%%%%%%%%%%%%%%%%%%%%%%%%%%%%%%%%%%%%%%%%%%%%%%%%%%

\section{\label{sec:gr}Input-output relations}

The equations of motion which follow from the Hamiltonian (\ref{2})
in Sec.~\ref{sec:ba.1} together with the Green tensor as
given in Sec.~\ref{sec:ba.2} can be used to study
various effects of the atom-light interaction in the presence of
dispersing and absorbing multilayered planar
structures (see the examples mentioned in Sec.~\ref{sec1}). 
Here, we restrict our attention to the general relation
between the fields outside and inside a multilayer plate
regardless of the generation of the fields by the 
active light sources. The theory thus includes the general
case, where sources are present in the both regions, outside
and inside the plate.

If the incoming fields, incident on the two boundary planes
of the plate (i.e., $z$ $\!=$ $\!0^-$ for \mbox{$j$ $\!=$ $\!0$}
and $z$ $\!=$ $\!0^+$ for \mbox{$j$ $\!=$ $\!n$}; cf. Fig.~\ref{fig}),
are known as well as the fields generated inside the plate,
one can calculate the fields outgoing
from the two boundary planes  by means of input-output relations.
Note that any two planes $z$ $\!=$ $\!z^{(0)}$ $\!\leq$ $0^-$
and $z$ $\!=$ $\!z^{(n)}$ $\!\ge$ $0^+$ for \mbox{$j$ $\!=$ $\!0$}
and \mbox{$j$ $\!=$ $\!n$}, respectively, also can be used
in principle. As we shall see, these input-output relations
are valid for both passive and active devices,
for arbitrary layer materials and arbitrary media surrounding
the plate and for both propagating-field components
 and evanescent-field components.

%%%%%%%%%%%%%%%%%%%%%%%%%%%%%%%%%%%%%%%%%%%%%%%%%%%%%%%%%%%%%%%%%%%%%%%%

\subsection{Input-output relations in the ${\bf k  }$-space}
\label{sec:gr.1}

Here and in the following we restrict our attention to
the electric field, noting that the corresponding
expressions for the magnetic field can readily be obtained
by applying Eq.~(\ref{8}).
Following the line suggested in Ref.~\cite{Gruner54}
and substituting the Green tensor (\ref{19})
for \mbox{$j$ $\!=$ $\!0$} and \mbox{$j$ $\!=$ $\!n$}
into Eq.~(\ref{16}), we may decompose the field operators
$\hat{\bf E}  ^{(0)}(z, {\bf k  }, \omega )$ and
$\hat{\bf E}  ^{(n)}(z, {\bf k  }, \omega )$
in the form of 
\begin{eqnarray}
      \label{29}
\lefteqn{
      \hat{\bf E}   ^{(0)}(z,{\bf k  }, \omega )
      = \sum _{q=p,s} \left[{\bf e}^{(0)}_{q+} ({\bf k})
      \hat{\mathrm{E}}_{q\,{\rm in}} ^{(0)} (z,{\bf k  }, \omega)
      \right.
} \nonumber\\*[1ex]
      &&\hspace{15ex}
      \left.
      +\, {\bf e}^{(0)}_{q-} ({\bf k  })
      \hat{\mathrm{E}}_{q\,{\rm out}} ^{(0)} (z,{\bf k  }, \omega)\right],
\end{eqnarray}
\begin{eqnarray}
      \label{30}
\lefteqn{
      \hat{\bf E}   ^{(n)}(z,{\bf k  }, \omega )
      = \sum _{q=p,s} \left[ {\bf e}^{(n)}_{q-} ({\bf k})
      \hat{\mathrm{E}}_{q\,{\rm in}}^{(n)}(z,{\bf k  }, \omega)
      \right.
} \nonumber\\[1ex]
      &&\hspace{15ex}
      \left.
      +\, {\bf e}^{(n)}_{q+} ({\bf k  })
      \hat{\mathrm{E}}_{q\,{\rm out}}^{(n)} (z,{\bf k}, \omega)\right].
      \end{eqnarray}
Here, the operators
\begin{eqnarray}
      \label{31}
\lefteqn{
      \hat{\mathrm{E}}_{q\,{\rm in}} ^{(0)} (z, {\bf k  }, \omega)
      = - \frac {\mu _0 \omega} {2 \beta _0}  e^{ i \beta _0 z}
}\nonumber \\ [1ex]
      &&\times\;
      \int _{-\infty} ^{z}\! \! \D z'\, e^{- i \beta _0 z'}
      \,\hat {\bf j} ^{(0)} (z', {\bf k }, \omega)
      \cdot {\bf e}^{(0)}_{q+}({\bf k}),  
\\[1ex]
      \label{32}
\lefteqn{
      \hat{\mathrm{E}}_{q\,{\rm in}}^{(n)} (z, {\bf k  }, \omega)
      = - \frac {\mu _0 \omega} {2 \beta _n} e^{- i \beta _n z}
}\nonumber \\[1ex]
      &&\times\;
      \int _{z} ^{\infty}  \D z'\,
      e^{ i \beta _n z'}
      \,\hat {\bf j} ^{(n)}(z', {\bf k  }, \omega)
      \cdot  {\bf e}^{(n)}_{q-}({\bf k}) 
      \end{eqnarray}
and      
\begin{eqnarray}
      \label{33}
\lefteqn{
      \hat{\mathrm{E}}_{q\,{\rm out}} ^{(0)} (z,{\bf k  }, \omega)
      = e^{- i \beta _0 z}
      \hat{\mathrm{E}}_{q\,{\rm out}} ^{(0)} ({\bf k  }, \omega)
}\nonumber\\[1ex]
      &&
      +\,e^{ - i \beta _0 z}
      \int _{z} ^{0}  \D z'\,e^{ i \beta _0 z'}
      \,\hat {\bf j} ^{(0)} (z', {\bf k }, \omega) \cdot
      {\bf e}^{(0)}_{q-}({\bf k}), \quad
\\[1ex]
      \label{34}
\lefteqn{
      \hat{\mathrm{E}}_{q\,{\rm out}}^{(n)} (z,{\bf k}, \omega)
      = e^{ i \beta _n z}
      \hat{\mathrm{E}}_{q\,{\rm out}}^{(n)} ({\bf k}, \omega)
}\nonumber\\[1ex]
      &&
      +\,e^{ i \beta _n z}
      \int _{0} ^{z}  \D z'\,
      e^{- i \beta _n z'}
      \,\hat {\bf j} ^{(n)} (z', {\bf k }, \omega)\cdot
      {\bf e}^{(n)}_{q+}({\bf k}), \qquad
      \end{eqnarray}
respectively, play the role of input and output amplitude
operators, where the input-output relations  
\begin{widetext}
\begin{eqnarray}
      \label{35}
      \left(
      \begin{array}{c}
      \hat{\mathrm{E}}_{q\,{\rm out}} ^{(0)}({\bf k  }, \omega)\\[1ex]
      \hat{\mathrm{E}}_{q\,{\rm out}}^{(n)}({\bf k  }, \omega)
      \end{array} \right)
      =\left(
      \begin{array}{cc}
      r^q_{0/n}({\bf k},\omega) & t_{n/0} ^q({\bf k},\omega)\\[1ex]
      t_{0/n} ^q({\bf k},\omega) &  r ^q _{n/0}({\bf k},\omega)
      \end{array} \right)
      \left(
      \begin{array}{c}
      \hat{\mathrm{E}}_{q\,{\rm in}} ^{(0)}({\bf k  },  \omega) \\[1ex]
      \hat{\mathrm{E}}_{q\,{\rm in}}^{(n)}({\bf k  },  \omega)
      \end{array} \right)
      + \sum _{j = 1 } ^{n-1} \left(
      \begin{array}{cc}
      \phi _{q\, 0+} ^{(j)}({\bf k},\omega)
      & \phi _{q\, 0-} ^{(j)}({\bf k},\omega) \\[1ex]
      \phi _{q\, n+} ^{(j)}({\bf k},\omega)
      & \phi _{q\, n-} ^{(j)}({\bf k},\omega)
      \end{array} \right)
      \left(
      \begin{array}{c}
      \hat{\mathrm{E}}_{q+}^{(j)}({\bf k  }, \omega)\\[1ex]
      \hat{\mathrm{E}}_{q-}^{(j)}({\bf k  }, \omega)
      \end{array} \right)\!,\quad
      \end{eqnarray}
\end{widetext}
are valid (for the commutation relations, see the appendix).
They relate the output amplitude operators at the
boundary planes of the multilayer plate to the input amplitude
operators at the boundary planes,
\begin{eqnarray}
      \label{36}
      \hat{\mathrm{E}}_{q\,{\rm in,out}} ^{(0)} ({\bf k  }, \omega)
      &=&
      \left.
      \hat{\mathrm{E}}_{q\,{\rm in,out}} ^{(0)} (z, {\bf k  }, \omega)
      \right|_{z=0^-},
      \\[1ex]
      \label{37}
      \hat{\mathrm{E}}_{q\,{\rm in,out}}^{(n)} ({\bf k  }, \omega)
      &=&
      \left.
      \hat{\mathrm{E}}_{q\,{\rm in,out}}^{(n)} (z, {\bf k  }, \omega)
      \right|_{z=0^+},
      \end{eqnarray}
and the intraplate amplitude operators
\begin{eqnarray}
      \label{38}
\lefteqn{      
      \hat{\mathrm{E}}_{q\pm}^{(j)}({\bf k  }, \omega)
      = - \frac {\mu _0 \omega} {2 \beta_j }
}
\nonumber\\[1ex]
      &&\times\;
      \int _{0} ^{d_j}  \D z' \,
      e^{\mp i \beta _j z'}
      \,\hat {\bf j}^{(j)} (z', {\bf k  }, \omega)\cdot
      {\bf e}^{(j)} _{q\pm}({\bf k})
\end{eqnarray}
($j$ $\!=$ $\!1,2,\ldots,n-1$), which are associated with the
excitation inside the layers of the plate.
In Eq.~(\ref{35}), the $\phi$-coefficients read
\begin{eqnarray}
      \label{39}
      \phi _{q\, 0+} ^{(j)} \,=\,  \frac { t_{j/0} ^q
      e^{2i\beta _{j} d_{j}}}{D_{qj}} r ^q _{j/n},
      &\ &
      \phi _{q\, 0-} ^{(j)} \,=\,  \frac { t_{j/0} ^q  }{D_{qj}}\,,
      \end{eqnarray}
\begin{eqnarray}
      \label{40}
      \phi _{q\, n+} ^{(j)} \,=\,  \frac { t_{j/n} ^q
      e^{i\beta _{j} d_{j}}}{D_{qj}}\,,
      &\ &
      \phi _{q\, n-} ^{(j)} \,=\,  \frac { t_{j/n} ^q
      e^{i\beta _{j} d_{j}} }{D_{qj}}r ^q _{j/0}.
      \quad
\end{eqnarray}
It should be pointed out that the first term
on the right-hand side in
Eq.~(\ref{19}), which gives rise to a local contribution
to the electric field, has been omitted in Eqs.~(\ref{29})
and (\ref{30}). Though this contribution is irrelevant for the
incoming and outgoing fields, it must be included in the
overall field operator in general, even if there are
effectively no sources at the points of observations.

It is not difficult to prove that (similar to the
one-dimensional case \cite{Gruner54}) the $z$-dependent amplitude
operators (\ref{31}) -- (\ref{34})
obey quantum Langevin equations,
\begin{eqnarray}
      \label{41}
\lefteqn{
      \frac{\partial}{\partial z}
      \hat{\mathrm{E}}_{q\,{\rm in}} ^{(0)} (z, {\bf k  }, \omega)
      = i \beta _0 \,\hat{\mathrm{E}}_{q\,{\rm in}} ^{(0)} (z, {\bf k  },
      \omega)
}
\nonumber\\* [1ex]
      &&\hspace{10ex}
      -\,\frac {\mu _0 \omega} {2 \beta _0}
      \,\hat {\bf j} ^{(0)} (z, {\bf k }, \omega)\cdot
      {\bf e}^{(0)}_{q+}({\bf k}), 
      \end{eqnarray}
\begin{eqnarray}
      \label{42}
\lefteqn{
      \frac{\partial}{\partial z}
      \hat{\mathrm{E}}_{q\,{\rm out}} ^{(0)} (z, {\bf k  }, \omega)
      = -i \beta _0 \,\hat{\mathrm{E}}_{q\,{\rm out}} ^{(0)}
      (z, {\bf k  },\omega)
}
\nonumber\\* [1ex]
      &&\hspace{10ex}
      +\,\frac {\mu _0 \omega} {2 \beta _0}
      \,\hat {\bf j} ^{(0)} (z, {\bf k }, \omega)\cdot
      {\bf e}^{(0)}_{q-}({\bf k}), 
      \end{eqnarray}
and similar equations are valid for
$\hat{\mathrm{E}}_{q\,{\rm in}} ^{(n)} (z, {\bf k  }, \omega)$
and $\hat{\mathrm{E}}_{q\,{\rm out}} ^{(n)} (z, {\bf k  }, \omega)$.
These equations together with Eq.~(\ref{35})
render it possible to easily calculate the
%output amplitude operators
input and output fields 
at any position outside the plate. Needless to say that  
other than the boundary planes \mbox{$z$ $\!=$ $0^-$}
and \mbox{$z$ $\!=$ $0^+$} of the plate can be chosen
as reference planes for formulating the input-output
relations. 

Equation (\ref{35}) represents the basic input-output relations
for the amplitude  operators in the Fourier space.
The coefficients therein are determined only by the (complex)
permittivities and thicknesses of the layers of the plate and the
permittivities of the surrounding media. In particular,
for \mbox{$\varepsilon''_{0,n}$ $\!\to$
$\!0$} (i.e., \mbox{$\varepsilon_{0,n}$ $\!\to$
$\!1$}) Eq.~(\ref{35}) immediately yields the
input-output relations for the special case of the plate
being surrounded by vacuum. 

The  input-output relations
in the form of Eq.~(\ref{35}) are generally valid,
independent of the mechanism of creation of the incoming
fields and the fields inside the layers of the plate.
It is worth noting that they take into account both
propagating waves and evanescent waves.
In particular in the case when the plate is surrounded
by vacuum, then the input and output
amplitude operators are associated with propagating waves
\mbox{($\omega/c$ $\!>$ $\!k$)} or evanescent waves
\mbox{($\omega/c$ $\!<$ $\!k$)} in $z$-direction.

The temporal evolution (in the Heisenberg picture) of the
amplitude operators is determined by the time dependence of
the basic variables $\hat{\bf f}({\bf r},\omega)$
[or equivalently, $\hat{\bf f}(z,{\bf k},\omega)$],
which is governed by the Hamiltonian (\ref{2}).
In the special case when the plate is free of active atomic
sources, then the fields inside the layers of the plate
represent Langevin noise sources associated with
material absorption.
If absorption is disregarded
and the plate is surrounded by vacuum that may contain
active sources, then Eq.~(\ref{35}) solves
a scattering problem of the type considered in Ref.~\cite{Carminati00},
namely scattering of fields containing evanescent   
components. In general, 
absorption cannot be disregarded and 
both atomic sources and Langevin noise sources
contribute to a field (e.g., in cavity QED), which can
contain both propagating and evanescent components
(e.g., in near-field scanning probe microscopy). 

The input-output relations (\ref{35}) enable one to calculate
correlation functions of the output field amplitudes
in terms of those of the input field amplitudes
and the amplitudes of the fields inside the plate. The simplest case
is the calculation of the expectation values of the field amplitudes.  
For example, in a typical scattering arrangement
the active light sources are located
outside the plate, so that the field inside the plate is
the absorption-assisted
random field whose thermal-equilibrium expectation value
vanishes. Application of Eq.~(\ref{35}) thus leads to
the expectation-value relations
\begin{equation}
      \label{43}
      \bigl\langle\hat{\mathrm{E}}_{q\,{\rm out}} ^{(0)}
      ({\bf k  }, \omega)\bigr\rangle
      = r ^q _{0/n}
      \bigl\langle\hat{\mathrm{E}}_{q\,{\rm in}} ^{(0)}
      ({\bf k  }, \omega)\bigr\rangle
      + t_{n/0}^q
      \bigl\langle\hat{\mathrm{E}}_{q\,{\rm in}}^{(n)}
      ({\bf k  }, \omega)\bigr\rangle,
\end{equation}
\begin{equation}
      \label{44}  
      \bigl\langle\hat{\mathrm{E}}_{q\,{\rm out}}^{(n)}
      ({\bf k  }, \omega)\bigr\rangle
      = t_{0/n}^q 
      \bigl\langle\hat{\mathrm{E}}_{q\,{\rm in}} ^{(0)}
      ({\bf k  }, \omega)\bigr\rangle      
      + r ^q _{n/0}
      \bigl\langle\hat{\mathrm{E}}_{q\,{\rm in}}^{(n)}
      ({\bf k  }, \omega)\bigr\rangle,
\end{equation}
which exactly correspond to 
standard results in classical optics.
On the other hand, when the active sources are located
(in a cavity-like system) inside the plate, we find that 
%\begin{widetext}
%\begin{eqnarray}
\begin{equation}
\label{45}
\begin{split}
%      &&
      &\bigl\langle\hat{\mathrm{E}}_{q\,{\rm out}} ^{(0)}
      ({\bf k  }, \omega)\bigr\rangle
      =
      \sum _ {j = 1}^{n-1}
      \frac{t_{j/0}^q}{D_{qj}}\,
      e ^{i \beta_{j} d_{j}}\\
      &\times\left[
      e ^{-i \beta_{j} d_{j}}
      \bigl\langle\hat{\mathrm{E}}_{q-}^{(j)}
      ({\bf k  }, \omega)\bigr\rangle
      + r^q _{j/n} e ^{i \beta_j d_j}
      \bigl\langle\hat{\mathrm{E}}_{q+}^{(j)}
      ({\bf k  }, \omega)\bigr\rangle
      \right],
\end{split}
\end{equation}
\begin{equation}
\label{46}
\begin{split}
%      \\[1ex]
%      &&
      \bigl\langle\hat{\mathrm{E}}_{q\,{\rm out}}^{(n)}
      &({\bf k  }, \omega)\bigr\rangle
      =
      \sum _ {j = 1}^{n-1}
      \frac{t_{j/n}^q}{D_{qj}}\,
      e ^{i \beta_j d_j}\\
      &\times\left[
      \bigl\langle\hat{\mathrm{E}}_{q+}^{(j)}({\bf k  }, \omega)\bigr\rangle
      + r^q _{j/0}
      \bigl\langle\hat{\mathrm{E}}_{q-}^{(j)}({\bf k  }, \omega)\bigr\rangle
      \right].
\qquad \;	
\end{split}
\end{equation}      
%\end{eqnarray}
%\end{widetext}
Needless to say that when there are no active sources
inside the plate, then \mbox{$\bigl\langle\hat{\mathrm{E}}_{q\pm}^{(j)}
({\bf k  }, \omega)\bigr\rangle$ $\!=$ $\!0$}
\mbox{($j$ $\!=$ $\!1,2,\ldots,n$ $\!-$ $\!1$)} is valid and
thus \mbox{$\bigl\langle\hat{\mathrm{E}}_{q\,{\rm out}}^{(0,n)}
({\bf k  }, \omega)\bigr\rangle$ $\!=$ $\!0$}.
Clearly, higher-order correlation functions of the outgoing
amplitude operators do not vanish in this case in general.
For example, 
the spectral intensity (in the ${\bf k}$-space)
of the radiation outgoing from a plane of
a plate in thermal
equilibrium at temperature $T$,
for chosen polarization, is proportional to
$w_{q\,{\rm out}}^{(0,n)}({\bf k},\omega)$, where
\begin{eqnarray}
\label{46a}
\lefteqn{
\bigl\langle \hat{\mathrm{E}}_{q\,{\rm out}} ^{(0,n)\, \dagger}
({\bf k  }, \omega)\,\hat{\mathrm{E}}_{q\,{\rm out}} ^{(0,n)}
({\bf k  }', \omega')\bigr\rangle
}
\nonumber\\[1ex]&&
= w_{q\,{\rm out}}^{(0,n)}({\bf k},\omega)\,\delta(\omega-\omega')
\delta({\bf k}-{\bf k}').
\end{eqnarray}
Applying Eq.~(\ref{35}) together with Eqs.~(\ref{38})
and (\ref{17}) and making use of
\begin{eqnarray}
\label{46b}
\lefteqn{
\bigl\langle \hat{f}_\mu^{(j)\,\dagger}(z,{\bf k},\omega)
\hat{f}_{\mu'}^{(j)}(z',{\bf k}',\omega')
\bigr\rangle
}
\nonumber\\[1ex]&&
= n(\omega,T)\,
\delta_{\mu\mu'} \delta(\omega-\omega')\delta({\bf k}-{\bf k}')
\end{eqnarray}
($j$ $\!\neq$ $\!0,n$), where
\begin{equation}
\label{46c}
n(\omega,T)
= \left[\exp\!\left(\frac{\hbar\omega}{k_{\rm B}T}\right) -1
\right]^{-1}
\end{equation}
($k_{\rm B}$, Boltzmann constant)
is the well-known Bose-Einstein distribution function,
we derive
\begin{eqnarray}
\label{46d}
\lefteqn{
w_{q\,{\rm out}}^{(n)}({\bf k},\omega) = n(\omega,T)
\sum_{j=1}^{n-1} \left|\frac{t^q_{j/n}}{D_{qj}}\right|^2
e^{-2\beta_j''d_j}
\left\{c_{q++}^{(j)}({\bf k},\omega)\right.
}
\nonumber\\[1ex]&&
\left.
+\, \bigl|r_{j/0}^q\bigr|^2 c_{q--}^{(j)}({\bf k},\omega)
+ \left[r_{j/0}^q c_{q-+}^{(j)}({\bf k},\omega)
+ {\rm c.c.}\right]\right\}\qquad
\end{eqnarray}
(and $w_{q\,{\rm out}}^{(0)}({\bf k},\omega)$ accordingly),
with the coefficients $c_{q\lambda\lambda'}^{(j)}({\bf k},\omega)$
\mbox{($\lambda,\lambda'$ $\!=$ $\!\pm$)}
being given in Eqs.~(\ref{A15}) and (\ref{A16}) in the appendix.

%%%%%%%%%%%%%%%%%%%%%%%%%%%%%%%%%%%%%%%%%%%%%%%%%%%%%%%%%%%%%%%%%%%%%%%%%%
\subsection{Input-output relations in the ${\bm \rho }$-space}
\label{sec:gr.2}

%^%%%%%%%%%%%%%%%%%%%%%%%%%%%%%%%%%%%%%%%%%%%%%%%%%%%%%%%%%%%%%%%%%%%%%%%%
The input-output relations (\ref{35}) in the ${\bf k }$-space can
be transformed, according to Eq.~(\ref{15}),
into the ${\bm \rho }$-space in a straightforward manner.
Writing the electric-field operator
\mbox{$ \hat{\bf E}   ^{(0)}({\bf r  }, \omega )$ $\!\equiv$
$\!\hat{\bf E}  ^{(0)}(z, {\bm \rho  }, \omega )$} at the boundaries
of the plate as
\begin{eqnarray}
      \label{47}
      \hat{\bf E}   ^{(0)}({\bm \rho  }, \omega )
      &=& \left.\hat{\bf E}  ^{(0)}(z, {\bm \rho  }, \omega )\right|_{z=0^-}
      \nonumber\\[1ex]
      &=&
      \hat{\bf{E}}_{q\,{\rm in}} ^{(0)} ({\bm \rho  }, \omega)
      +
      \hat{\bf{E}}_{q\,{\rm out}} ^{(0)} ({\bm \rho  }, \omega) ,
     \\[1ex]
\nonumber\\      
      \label{48}
      \hat{\bf E}   ^{(n)}({\bm \rho  }, \omega )
      &=& \left.\hat{\bf E}  ^{(n)}(z, {\bm \rho  }, \omega )\right|_{z=0^+}
      \nonumber\\*[1ex]
      &=&\ 
      \hat{\bf{E}}_{q\,{\rm in}}^{(n)}({\bm \rho  }, \omega)
      +
     \hat{\bf{E}}_{q\,{\rm out}}^{(n)} ({\bm \rho}, \omega) ,
      \end{eqnarray}
we derive
\begin{widetext}
\begin{eqnarray}
\label{49}
\lefteqn{
   \hat{\bf E} _{{\rm out}}  ^{(0)}({\bm \rho  }, \omega )
   = \int \D^2 {\rho}'\,
   \mbb{R}_{0/n} ({\bm \rho}, {\bm \rho '}, \omega)\cdot
   \hat{\bf E} _{{\rm in}}  ^{(0)}({\bm \rho '}, \omega )
   + \int \D^2 {\rho}'\,
   \mbb{T}_{n/0} ({\bm \rho}, {\bm \rho '}, \omega)\cdot
   \hat{\bf E} _{{\rm in}}  ^{(n)}({\bm \rho '}, \omega )
}\nonumber\\[1ex]&&\hspace{10ex}
+ \sum _{j=1} ^{n-1} \int  \D^2 {\rho}'\,
	\left\lbrace \mbb{\Phi} _{ 0+} ^{(j)}
        ({\bm \rho}, {\bm \rho '}, \omega)\cdot
   \hat{\bf E} _{+}  ^{(j)}({\bm \rho '}, \omega )
+  \mbb{\Phi} _{ 0-} ^{(j)} ({\bm \rho}, {\bm \rho '}, \omega)\cdot
   \hat{\bf E} _{-}  ^{(j)}({\bm \rho '}, \omega )\right\rbrace,
\\[1ex]
\label{50}
\lefteqn{
   \hat{\bf E} _{{\rm out}}  ^{(n)}({\bm \rho  }, \omega )
   = \int \D^2 {\rho}'\,
   \mbb{R}_{n/0} ({\bm \rho}, {\bm \rho '}, \omega)\cdot
   \hat{\bf E} _{{\rm in}}  ^{(n)}({\bm \rho '}, \omega )
   + \int \D^2 {\rho}'\,
   \mbb{T}_{0/n} ({\bm \rho}, {\bm \rho '}, \omega)\cdot
   \hat{\bf E} _{{\rm in}}  ^{(0)}({\bm \rho '}, \omega )
}\nonumber\\[1ex]&&\hspace{10ex}
+ \sum _{j=1} ^{n-1} \int  \D^2 {\rho}'\,
	\left\lbrace \mbb{\Phi} _{ n+} ^{(j)}
        ({\bm \rho}, {\bm \rho '}, \omega)\cdot
   \hat{\bf E} _{+}  ^{(j)}({\bm \rho '}, \omega )
+  \mbb{\Phi} _{ n-} ^{(j)} ({\bm \rho}, {\bm \rho '}, \omega)\cdot
   \hat{\bf E} _{-}  ^{(j)}({\bm \rho '}, \omega )\right\rbrace,
\end{eqnarray}
\end{widetext}
where the electric-field operators
$\hat{\bf E} _{{\rm in}}  ^{(0,n)}({\bm \rho  }, \omega )$
and $\hat{\bf E} _{{\rm \pm}}  ^{(j)}({\bm \rho  }, \omega )$
are respectively related to the amplitude operators 
$\hat{E}_{q\,{\rm in}}^{(0,n)}({\bf k},\omega)$
and $\hat{E}_{q\pm}^{(j)}({\bf k},\omega)$
\mbox{($j$ $\!=$ $\!1,2,\ldots,n$ $\!-$ $\!1$)} as
\begin{eqnarray}
\label{51}
   \hat{\bf E} _{{\rm in}}  ^{(0)}({\bm \rho  }, \omega )
   \!=\! \frac {1} {(2 \pi )^2}\!\int\! \D^2 {k}\sum _{q=p,s}\!
   {\bf e}^{(0)}_{q+} ({\bf k})
   \hat{\mathrm{E}}_{q\,{\rm in}} ^{(0)} ({\bf k  }, \omega)
   e^{i{\bf k } {\bm {\rho}} }\! , \quad \,
\\[1ex]
\label{52}
   \hat{\bf E} _{{\rm in}}  ^{(n)}({\bm \rho  }, \omega )
   \!=\! \frac {1} {(2 \pi )^2}\!\int \!\D^2 {k}\sum _{q=p,s}\!
   {\bf e}^{(n)}_{q-} ({\bf k})
   \hat{\mathrm{E}}_{q\,{\rm in}}^{(n)} ({\bf k  }, \omega)
   e^{i{\bf k } {\bm {\rho}} }\!, \quad
\\[1ex]
\label{53}
   \hat{\bf E} _{{\rm \pm}}  ^{(j)}({\bm \rho  }, \omega )
   \!=\! \frac {1} {(2 \pi )^2}\!\int \!\D^2 {k}\sum _{q=p,s}\!
   {\bf e}^{(j)}_{q\pm} ({\bf k})
   \hat{\mathrm{E}}_{q\,{\rm \pm}}^{(j)} ({\bf k  }, \omega)
   e^{i{\bf k } {\bm {\rho}} }\!. \quad \;
\end{eqnarray}
In the above,
the tensor-valued reflection and transmission
integral kernels, respectively, read as
\begin{eqnarray}
\label{54}
\lefteqn{
      {\bf R}_{0/n} ({\bm \rho}, {\bm \rho '}, \omega)
}
\nonumber\\&&
      =
      \frac {1} {(2 \pi)^2}\int \D^2 {k}\sum _{q=p,s}
      {\bf e}^{(0)}_{q-} ({\bf k  }) r^q _{0/n}
      {\bf e}^{(0)}_{q+}({\bf k}) e^{i{\bf k}
      ({\bm \rho}-{\bm \rho'})},\qquad
\\[1ex]
\label{55}
\lefteqn{
      {\bf R}_{n/0} ({\bm \rho}, {\bm \rho '}, \omega)}
\nonumber \\&&
      = \frac {1} {(2 \pi)^2}\int \D^2 {k}\sum _{q=p,s}
      {\bf{e}}^{(n)}_{q+} ({\bf k  }) r^q _{n/0}{\bf{e}}^{(n)}_{q-}
      ({\bf k})
      e^{i{\bf k} ({\bm \rho}-{\bm \rho'})}\qquad
     \end{eqnarray}
and
\begin{eqnarray}
\label{56}
\lefteqn{
      {\bf T}_{0/n} ({\bm \rho}, {\bm \rho '}, \omega) 
}
\nonumber \\&&
      = \frac {1} {(2 \pi)^2}\int \D^2 {k} \sum _{q=p,s}
      {\bf e }^{(n)}_{q+} ({\bf k  }) t^q _{0/n}{\bf e}^{(0)}_{q+}
      ({\bf k})
      e^{i{\bf k} ({\bm \rho}-{\bm \rho'})},\qquad
\\[1ex]
\label{57}
\lefteqn{
      {\bf T}_{n/0} ({\bm \rho}, {\bm \rho '}, \omega)
}
\nonumber \\&&
      = \frac {1} {(2 \pi)^2}\int \D^2 {k} \sum _{q=p,s}
      {\bf{e}}^{(0)}_{q-} ({\bf k  }) t^q _{n/0}{\bf{e}}^{(n)}_{q-}({\bf k})
      e^{i{\bf k} ({\bm \rho}-{\bm \rho'})},\qquad
\end{eqnarray}
and the integral kernels that relate the outgoing fields to
the intraplate fields are given by
\begin{eqnarray}
\lefteqn{
\label{58}
      {\bf \Phi }_{0\pm}^{(j)} ({\bm \rho}, {\bm \rho '}, \omega)
}
\nonumber \\&&
      = \frac {1} {(2 \pi)^2}\int \D^2 {k}\sum _{q=p,s}
      {\bf e}^{(0)}_{q-} ({\bf k  }) \phi ^{(j)} _{q \,0 \pm }
      {\bf{e}}^{(j)}_{q\pm}({\bf k})
      e^{i{\bf k} ({\bm \rho}-{\bm \rho'})},\qquad
\\[1ex]
\lefteqn{
\label{59}
      {\bf \Phi}_{n\pm} ({\bm \rho}, {\bm \rho '}, \omega) 
}
\nonumber \\&&
      = \frac {1} {(2 \pi)^2}\int \D^2 {k} \sum _{q=p,s}
      {\bf e}^{(j)}_{q+} ({\bf k  }) \phi^{(j)} _{q\, n \pm }
      {\bf e}^{(j)}_{q\pm}({\bf k})
      e^{i{\bf k} ({\bm \rho}-{\bm \rho'})}.\qquad
\end{eqnarray}
Note that the integral kernels depend only on the difference
\mbox{${\bm \rho}$ $\!-$ $\!{\bm \rho'}$}, because of
the translational symmetry in the $x$- and $y$-directions.
The input-output relations in the form of
Eqs.~(\ref{49}), (\ref{50}) can be considered as a special
realization of the general relations given
for a dielectric scattering region of arbitrary shape
in Ref.~\cite{Scheel3D}.

%%%%%%%%%%%%%%%%%%%%%%%%%%%%%%%%%%%%%%%%%%%%%%%%%%%%%%%%%%%%%%%%
\subsection{Input-output relations in terms of bosonic
operators in the {\bf k}-space}
\label{sec:gr.3}
%%%%%%%%%%%%%%%%%%%%%%%%%%%%%%%%%%%%%%%%%%%%%%%%%%%%%%%%%%%%%%%%

\subsubsection{General case}
\label{general}

In many applications it may be advantageously to express the incoming
and the outgoing field operators in  terms of
appropriately chosen bosonic operators.
Recalling the basic commutation relations (\ref{4}) and
(\ref{5}) and the orthogonality of polarization unit vectors
as given by Eqs.~(\ref{27}) and (\ref{28}),
from Eqs.~(\ref{31}) and (\ref{32}) together with
Eq.~(\ref{17}) and (\ref{18}) and Eqs.~(\ref{36}) and (\ref{37})
we find that the input amplitude operators satisfy the
commutation relations
\begin{eqnarray}
      \label{60}
\lefteqn{
      \left[\hat{\mathrm{E}}_{q\,{\rm in}}^{(0,n)}
      ({\bf k  }, \omega),
      \hat{\mathrm{E}}_{q'\,{\rm in}}^{(0,n)\dagger}
      ({\bf k  }', \omega')\right]
}
\nonumber \\[1ex]&&
      \,=\, c_{q\,{\rm in}}^{(0,n)}({\bf k},\omega)\,
      \delta_{qq'}\delta(\omega-\omega')\delta({\bf k}-{\bf k}'),
\end{eqnarray}
where
\begin{eqnarray}
      \label{61}
      c_{q\,{\rm in}}^{(0)} 
      = \frac{\pi \hbar}{\epsilon_0}
     \left(\frac{\omega}{c}\right)^2
      \frac{\beta_{0}'}{|\beta_{0}|^2}
      \,{\bf e}^{(0)}_{q+} ({\bf k  })\cdot{\bf e}^{(0)*}_{q+}
      ({\bf  k}),
\\[1ex]
      \label{62}
      c_{q\,{\rm in}}^{(n)} 
      = \frac{\pi \hbar}{\epsilon_0}
     \left(\frac{\omega}{c}\right)^2
      \frac{\beta_{n}'}{|\beta_{n}|^2}
      \,{\bf e}^{(n)}_{q-} ({\bf k  })\cdot{\bf e}^{(n)*}_{q-}
      ({\bf  k}).
      \end{eqnarray}
Similarly, the output amplitude operators and the intraplate
amplitude operators can be shown to satisfy commutation
relations of the same type, i.e.,
%\mbox{($\lambda$ $\!=$ $\!\pm$)}  
\begin{eqnarray} 
      \label{63}
\lefteqn{
      \left[\hat{\mathrm{E}}_{q\,{\rm out}}^{(0,n)}
      ({\bf k  }, \omega),
      \hat{\mathrm{E}}_{q'\,{\rm out}}^{(0,n)\dagger}
      ({\bf k  }', \omega')\right]
}
\nonumber \\[1ex]&&
      =\, c_{q\,{\rm out}}^{(0,n)}({\bf k},\omega)\,
      \delta_{qq'}\delta(\omega-\omega')\delta({\bf k}-{\bf k}'),
\\[1ex]
      \label{64}
\lefteqn{
      \left[\hat{\mathrm{E}}_{q\lambda}^{(j)}
      ({\bf k  }, \omega),
      \hat{\mathrm{E}}_{q'\lambda'}^{(j)\dagger}
      ({\bf k  }', \omega')\right]}
\nonumber \\[1ex]&&
       =\,c_{q\lambda\lambda'}^{(j)}({\bf k},\omega)\,
       \delta_{qq'}\delta(\omega-\omega')\delta({\bf k}-{\bf k}')
\end{eqnarray}
[for the coefficients $c_{q\,{\rm out}}^{(0,n)}({\bf k},\omega)$
and $c_{q\lambda\lambda'}^{(j)}({\bf k},\omega)$, respectively,
see Eqs.~(\ref{A12}), (\ref{A13}) and
Eqs.~(\ref{A15}), (\ref{A16}) in the appendix].
Needless to say that input amplitude operators, that
refer to different sides of the plate, commute, 
input amplitude operators and intraplate amplitude
operators commute, and intraplate amplitude operators that refer
to different layers also commute. Note that 
output amplitude operators that refer to different sides of
the plate do not commute in general
[see Eqs.~(\ref{A9}) and (\ref{A10}) in the appendix],
which is similar as in the one-dimensional situation
(cf. \cite{Gruner54}).

From Eqs.~(\ref{60}) and (\ref{63}) it is seen that
bosonic input and output operators can be
introduced according to 
\begin{equation}
      \label{65}
      \hat{\mathrm{E}}_{q\,{\rm in,out}}^{(0,n)} ({\bf k  }, \omega)
      = \sqrt{c_{q\,{\rm in,out}}^{(0,n)}({\bf k},\omega)} \, 
      \hat{a}_{q\,{\rm in,out}}^{(0,n)}({\bf k},\omega),
\end{equation}
thus
\begin{eqnarray}
      \label{66}
\lefteqn{
      \left[\hat{a}_{q\,{\rm in,out}}^{(0,n)} ({\bf k  }, \omega),
      \hat{a}_{q'\,{\rm in,out}}^{(0,n)\dagger} ({\bf k  }', \omega')
      \right]
}
\nonumber \\[1ex]&&\hspace{6ex}
      =\, 
      \delta_{qq'}\delta(\omega-\omega')\delta({\bf k}-{\bf k}').
      \end{eqnarray}
Since for \mbox{$\lambda$ $\!\neq$ $\!\lambda'$}
in Eq.~(\ref{64}) \mbox{$c_{q\lambda\lambda'}^{(j)}({\bf k},\omega)$
$\!\neq$ $\!0$} is valid in general, it is useful to introduce
intraplate bosonic operators according to the superposition
\begin{equation}
      \label{67}
      \hat{\mathrm{E}}_{q\lambda}^{(j)} ({\bf k  }, \omega)
      = \sum_{\lambda'=\pm}
      \tau_{q\lambda\lambda'}^{(j)}({\bf k  }, \omega) \,
      \hat{a}_{q\lambda'}^{(j)} ({\bf k  }, \omega) 
\end{equation}
\mbox{($j$ $\!=$ $\!1,\ldots,n$ $\!-$ $\!1$)} and choosing
the coefficients $\tau_{q\lambda\lambda'}^{(j)}({\bf k  }, \omega)$
[Eqs.~(\ref{A20}) and (\ref{A21}) in the appendix]
in such a way that
\begin{eqnarray}
      \label{68}
\lefteqn{
      \left[\hat{a}_{q\lambda }^{(j)}  ({\bf k  }, \omega),
      \hat{a}_{q'\lambda'}^{(j)\dagger} ({\bf k ' }, \omega ') \right]
}
\nonumber\\&&\hspace{6ex}
      =\,\delta_{\lambda\lambda'}\delta_{qq'}
      \delta( \omega -\omega ') \delta({\bf k } - {\bf k '}) .
      \end{eqnarray}

Substituting Eqs.~(\ref{65}) and (\ref{67}) into
Eq.~(\ref{35}), we may express the input-output relations
in the ${\bf k}$-space in terms of bosonic operators,
\begin{widetext}
\begin{eqnarray}
      \label{69}
      \left(
      \begin{array}{c}
      \hat{a}_{q\,{\rm out}}^{(0)} ({\bf k  }, \omega)\\[1ex]
      \hat{a}_{q\,{\rm out}}^{(n)} ({\bf k  }, \omega)
      \end{array} \right)
      =\left(
      \begin{array}{cc}
      \tilde{r}^q_{0/n}({\bf k},\omega)
      & \tilde{t}_{n/0} ^q({\bf k},\omega)\\[1ex]
      \tilde{t}_{0/n} ^q({\bf k},\omega)
      &  \tilde{r} ^q _{n/0}({\bf k},\omega)
      \end{array} \right)
      \left(
      \begin{array}{c}
      \hat{a}_{q\,{\rm in}}^{(0)} ({\bf k  },  \omega) \\[1ex]
      \hat{a}_{q\,{\rm in}}^{(n)} ({\bf k  },  \omega) 
      \end{array} \right)
      + \sum _{j = 1 } ^{n-1} \left(
      \begin{array}{cc}
      \tilde{\phi} _{q\, 0+} ^{(j)}({\bf k},\omega)
      & \tilde{\phi} _{q\, 0-} ^{(j)}({\bf k},\omega) \\[1ex]
      \tilde{\phi} _{q\, n+} ^{(j)}({\bf k},\omega)
      & \tilde{\phi} _{q\, n-} ^{(j)}({\bf k},\omega)
      \end{array} \right)
      \left(
      \begin{array}{c}
      \hat {a}_{q+}^{(j)} ({\bf k  }, \omega)\\[1ex]
      \hat {a}_{q-}^{(j)} ({\bf k  }, \omega)
      \end{array} \right)\!,\quad
      \end{eqnarray}
\end{widetext}
where the coefficients [modified in comparison with Eq.~(\ref{35})] 
read
 
\begin{eqnarray}
      \label{70}
      \tilde{r}^q_{0/n} = 
      \sqrt{ \frac{c_{q\,{\rm in}}^{(0)} } 
      {c_{q\,{\rm out}}^{(0)} } }
      \,r ^q _{0/n} \,,  
      \qquad
      \tilde{t}_{n/0} ^q = 
      \sqrt{ \frac{c_{q\,{\rm in}}^{(n)} }
      {c_{q\,{\rm out}}^{(0)} }}
      \, t_{0/n} ^q \,,
\\[1ex]
      \label{71}
      \tilde{t} ^q _{0/n} =   
      \sqrt{ \frac{c_{q\,{\rm in}}^{(0)} }
      {c_{q\,{\rm out}}^{(n)}} }
      \,t ^q _{n/0} \,,
      \qquad
      \tilde{r}_{n/0}^q = 
      \sqrt{ \frac{c_{q\,{\rm in}}^{(n)} }
      {c_{q\,{\rm out}}^{(n)} } }
      \,r_{n/0} ^q \,,
\end{eqnarray}
and
\begin{equation}
      \label{nn09}
      \tilde{\phi}_{q\,0,n\,\lambda} ^{(j)}
      =  \left(c_{q\,{\rm out}}^{(0,n)}\right)^{-1/2} 
      \sum_{\lambda'=+,-} \phi_{q\,0,n\,\lambda'} ^{(j)} 
      \,\tau_{q\lambda'\lambda}^{(j)} \,.
\end{equation}
Note that for \mbox{${\bf k}$ $\!=$ $\!0$}
(i.e., perpendicular incidence of light on the plate),
the input and output relations (\ref{69}) exactly 
agree with previous results \cite{Gruner54} obtained
within a one-dimensional treatment.

%It should be pointed out that in the three-dimensional
%treatment the introduction of the
%bosonic operators makes only sense if the commutators
%of the correspondig amplitude operators [Eqs.~(\ref{60}), 
%(\ref{63}), and (\ref{64})] do not vanish,
%i.e., the coefficients $c_{q\,{\rm in,out}}^{(0,n)}$
%and $c_{q\lambda\lambda'}^{(j)}$ must not vanish.
%From Eqs.~(\ref{61}) and (\ref{62}) it is seen that
%$c_{q\,{\rm in}}^{(0)}$ and $c_{q\,{\rm in}}^{(n)}$,
%respectively, do not vanish, if $\beta_0'$ and
%$\beta_n'$ are different from zero. 
 
%-----------------------------------------------------------------------
 
\subsubsection{Plate surrounded by vacuum}
\label{vacuum}

Let us turn to the limiting case that the space outside
the plate -- except for possible active atomic sources -- may be
regarded as being vacuum, i.e.,
\begin{equation}
\label{73}
\varepsilon_{0,n}''(\omega) \to 0,
\quad
\varepsilon_{0,n}'(\omega) \to 1,
\end{equation} 
\begin{equation}
\label{74}
\beta_{0,n}'(\omega,k) \to
\left\{
\begin{array}{l@{\quad\mbox{if}\quad}l}
0 & \omega/c \leq k,\\[1ex]
\sqrt{\omega^2/c^2 - k^2} & \omega/c > k.
\end{array}
\right.
\end{equation}
 
%Only for
For
the propagating-field components observed for 
\mbox{$\omega/c$ $\!>$ $\!k$},
%bosonic input and output operators can be defined.
%In this case,
the coefficients $c_{q\,{\rm in}}^{(0,n)}$
and $c_{q\,{\rm out}}^{(0,n)}$, respectively, read 
\begin{equation}
      \label{75}
      c_{q\,{\rm in}}^{(0)} = c_{q\,{\rm in}}^{(n)}
      = \frac{\pi \hbar}{\epsilon _0}
      \left(\frac{\omega}{c}\right)^2
      \frac{1}{\beta_{0}}
\end{equation}
and
\begin{eqnarray}
      \label{76}
      c_{q\,{\rm out}}^{(0,n)} 
      \,=\ c_{q\,{\rm in}}^{(0,n)},
      \end{eqnarray}
%are obtained,
as it can be seen from Eqs.~(\ref{61}) and (\ref{62})
and Eqs.~(\ref{A23}) and (\ref{A24}) in the appendix.
Hence, bosonic  input and output operators can be defined
according to Eq.~(\ref{65}). Moreover, the output amplitude
operators that refer to different sides of the plate
commute [Eq.~(\ref{A22}) in
the appendix], which implies that the
associated bosonic operators commute as well.
Making use of Eqs.~(\ref{75}) and (\ref{76}), we see
that the modified reflection and transmission coefficients
(\ref{70}) and (\ref{71}), respectively, become identical
with the original ones.
For \mbox{$\omega/c$ $\!>$ $\!k$}, Eq.~(\ref{69}) thus reduces to
\begin{widetext}
\begin{eqnarray}
      \label{77}
      \left(
      \begin{array}{c}
      \hat{a}_{q\,{\rm out}}^{(0)} ({\bf k  }, \omega)\\[1ex]
      \hat{a}_{q\,{\rm out}}^{(n)} ({\bf k  }, \omega)
      \end{array} \right)
      =\Biggl(
      \begin{array}{cc}
      r^q_{0/n}({\bf k  }, \omega) & t_{n/0} ^q({\bf k  }, \omega)\\[1ex]
      t_{0/n} ^q({\bf k  }, \omega) &  r ^q _{n/0}({\bf k  }, \omega)
      \end{array} \Biggr)
      \left(
      \begin{array}{c}
      \hat{a}_{q\,{\rm in}}^{(0)} ({\bf k  },  \omega) \\[1ex]
      \hat{a}_{q\,{\rm in}}^{(n)} ({\bf k  },  \omega) 
      \end{array} \right)
      + \sum _{j = 1 } ^{n-1} \left(
      \begin{array}{cc}
      \tilde{\phi} _{q\, 0+} ^{(j)}({\bf k  }, \omega)
      & \tilde{\phi} _{q\, 0-} ^{(j)}({\bf k  }, \omega) \\[1ex]
      \tilde{\phi} _{q\, n+} ^{(j)}({\bf k  }, \omega)
      & \tilde{\phi} _{q\, n-} ^{(j)}({\bf k  }, \omega)
      \end{array} \right)
      \left(
      \begin{array}{c}
      \hat {a}_{q+}^{(j)} ({\bf k  }, \omega)\\[1ex]
      \hat {a}_{q-}^{(j)} ({\bf k  }, \omega)
      \end{array} \right),\quad
      \end{eqnarray}
      \end{widetext}
where the transformation matrix  connecting the bosonic
output operators with the bosonic input operators is exactly the same
as that for the corresponding amplitude operators
in Eq.~(\ref{35}), and the matrix equation
\begin{widetext}
\begin{eqnarray}
      \label{78}
      \left(
      \begin{array}{cc}
      r^q_{0/n} & t_{n/0} ^q\\[1ex]
      t_{0/n} ^q &  r ^q _{n/0}
      \end{array} \right)
      \left(
      \begin{array}{cc}
      r^{q\ast}_{0/n} & t_{0/n} ^{q\ast}\\[1ex]
      t_{n/0} ^{q\ast} &  r ^{q\ast} _{n/0}
      \end{array} \right)
      + \sum _{j = 1 } ^{n-1} \left(
      \begin{array}{cc}
      \tilde{\phi} _{q\, 0+} ^{(j)}
      & \tilde{\phi} _{q\, 0-} ^{(j)} \\[1ex]
      \tilde{\phi} _{q\, n+} ^{(j)}
      & \tilde{\phi} _{q\, n-} ^{(j)}
      \end{array} \right)
      \left(
      \begin{array}{cc}
      \tilde{\phi} _{q\, 0+} ^{(j)\ast}
      & \tilde{\phi} _{q\, n+} ^{(j)\ast} \\[1ex]
      \tilde{\phi} _{q\, 0-} ^{(j)\ast}
      & \tilde{\phi} _{q\, n-} ^{(j)\ast}
      \end{array} \right)
      = \left(
      \begin{array}{cc}
      1 & 0 \\[1ex]
      0 & 1 
      \end{array} \right)
      \end{eqnarray}
      \end{widetext}
is valid. In this case, the field operators
${\bf \hat{E}} ^{(0)}(z,{\bf k  }, \omega )$
and ${\bf \hat{E}}^{(n)}(z,{\bf k  }, \omega )$
at the boundaries of the multilayer plate
[see Eqs.~(\ref{29}), (\ref{30}), (\ref{36}), and (\ref{37})]
can be represented as 
\begin{widetext}
\begin{eqnarray}
 {\bf \hat{E}}   ^{(0)}(0^-,{\bf k  }, \omega ) =
 -\frac { \omega } {c } \sqrt{\frac {\pi \hbar } {\beta _0\epsilon _0 }}
 \sum _{q=p,s}\,
 \left[
 {\bf{e}}^{(0)}_{q+} ({\bf k}) \hat{a}_{q\,{\rm in}}^{(0)} ({\bf k  },
 \omega)  +
 {\bf{e}}^{(0)}_{q-} ({\bf k  })\hat{a}_{q\,{\rm out}}^{(0)}
 ({\bf k  }, \omega)\right] , 
\label{79}\\[1ex]
 {\bf \hat{E}}   ^{(n)}(0^+,{\bf k  }, \omega ) =
 -\frac { \omega } {c} \sqrt{\frac {\pi \hbar 
 } {\beta _n\epsilon _0 }}
 \sum _{q=p,s}\,
 \left[ {\bf{e}}_{q-}^{(n)} ({\bf k}) \hat{a}_{q\,{\rm in}}^{(n)}
 ({\bf k  }, \omega)  +
 {\bf{e}}_{q+}^{(n)} ({\bf k  })\hat{a}_{q\,{\rm out}}^{(n)}
 ({\bf k  }, \omega)\right]  
\label{80}
\end{eqnarray}
\end{widetext}
($\beta_0$ $\!=$ $\!\beta_n$ $\!=$ $\!\sqrt{\omega^2/c^2-k^2}$,
$\omega/c$ $\!>$ $\!k$).
Note that in Eqs.~(\ref{79}) and (\ref{80}) is nothing said about
the location of the active light sources. Clearly, if
there are no active light sources outside the plate, then
${\bf \hat{E}} ^{(0)}(z,{\bf k  }, \omega )$ 
and ${\bf \hat{E}}^{(n)}(z,{\bf k  }, \omega )$
evolve effectively freely there [cf. Eqs.~(\ref{33})
and (\ref{34})].  
 
%Thus, we find that for
For the evanescent-field components observed for
\mbox{$\omega/c$ $\!\leq$ $\!k$},
%that is to say, for the evanescent-field components,
the coefficients $c_{q\,{\rm in}}^{(0)}$ [Eq.~(\ref{61})]
and $c_{q\,{\rm in}}^{(n)}$ [Eq.~(\ref{62})] identically
vanish,
because of $\beta'_0$ $\!=$ $\!\beta'_n$ $\!=$ $\!0$.
Recalling Eq.~(\ref{65}), we see that  
%and
bosonic input operators cannot be introduced for
the evanescent-field components.
%-- a point which has been
%%overlooked
%disregarded
%in Ref.~\cite{Savasta19}.
Hence, it is impossible to extend the validity of Eq.~(\ref{77}) to
the evanescent-field components. 
%Needless to say that the
%original input-output relations (\ref{35}) of course apply.
To treat
%the evanescent-field components,
them,
one has therefore
to go back to the generally valid input-output relations
(\ref{35}) for the amplitude operators.
Clearly, when there are no active light sources 
at any finite distance from the plate, then
the input field can be regarded as being effectively
a propagating field [see Eqs.~(\ref{31}) and (\ref{32})],
that is to say, a field that does not contain evanescent
input components. However, the output field may contain
evanescent output components resulting from the intraplate
field.

%%%%%%%%%%%%%%%%%%%%%%%%%%%%%%%%%%%%%%%%%%%%%%%%%%%%%%%%%%%%%%%%%%%%%%%%%%%
%%%%%%%%%%%%%%%%%%%%%%%%%%%%%%%%%%%%%%%%%%%%%%%%%%%%%%%%%%%%%%%%%%%%%%%%%%%

\section{\label{sec:co}Summary and concluding remarks} 

Applying the Green-tensor formalism of the quantization
of the electromagnetic field in the presence of inhomogeneous,
causal, linear dielectric bodies,
we have derived the most general
three-dimensional quantum input-output relations 
for the electromagnetic-field operators at
an arbitrary multilayer dielectric plate,
and we have given the commutation relations needed.
Taking into account
(i) dispersion and absorption of the plate and the
surrounding medium,
(ii) possible active light sources inside and/or outside
the plate, and
(iii) both propagating-field components
and evanescent-field components, 
the input-output relations connect the output field
operators at the boundaries of the plate with
the input field operators at the boundaries
of the plate and the fields inside the layers
of the plate. The outgoing fields outside
the plate can then be obtained from the ones
at the boundaries of the plate via quantum Langevin
equations, with the Langevin noise sources being
determined by the active and passive light sources
in the respective half-space outside the plate.   

We have given the input-output relations in terms of both algebraic
relations for amplitude operators in the ${\bf k}$-space
and integral relations for field operators  
in the coordinate space. Further, we have shown that in the
${\bf k}$-space it is possible, provided that the incoming
fields contain effectively propagating components, 
to rewrite the input-output relations in order to obtain them in
terms of relations between bosonic operators instead of
amplitude operators. Therefore the input-output relations for the
amplitude operators are more general than those for the bosonic operators.

Finally, we have studied the limiting case of the plate being
surrounded by vacuum. In this case, the
input-output relations in terms of bosonic operators only
apply to the propagating-field components, for which
\mbox{$\omega/c$ $\!>$ $\!k$} is valid. An application
to the evanescent-field components
\mbox{($\omega/c$ $\!\leq$ $\!k$)}
would run into contradictions.
To treat them, one has therefore to go back to the amplitude operators.

The derived input-output relations generalize previous results
obtained from a one-dimensional calculation \cite{Gruner54}).
In the special case of a plate that is embedded in vacuum,
without active light sources at any finite distance from
the plate (and without active light inside the plate),
they reduce to those given in Ref.~\cite{Savasta19},
if evanescent-field components are disregarded.

In conclusion, the derived input-output relations are suited for
studying the quantum statistical properties of electromagnetic
fields in the presence of planar multilayered structures
(such as cavity-like systems or photonic crystals),
with special emphasis on absorption-assisted quantum decoherence,
including evanescent-field effects.

\begin{acknowledgments}
We would like to thank Ho Trung Dung and C. Raabe for useful discussions.

\end{acknowledgments}

\appendix*
%%%%%%%%%%%%%%%%%%%%%%%%%%%%%%%%%%%%%%%%%%%%%%%%%%%%%%%%%%%%%%%
\section{\label{app}Commutation relations}
%%%%%%%%%%%%%%%%%%%%%%%%%%%%%%%%%%%%%%%%%%%%%%%%%%%%%%%%%%%%%%%

Let us  begin with the calculation of the
commutators of the output amplitude operators 
$\hat{\mathrm{E}}_{q\,{\rm out}} ^{(0,n)} ({\bf k  }, \omega)$.
From Eq.~(\ref{35}) it follows that
\begin{eqnarray}
      \label{A1}
\lefteqn{
      \hat{\mathrm{E}}_{q\,{\rm out}} ^{(0)} ({\bf k  }, \omega)
      =\, r^q_{0/n} \hat{\mathrm{E}}_{q\,{\rm in}} ^{(0)}
      ({\bf k  },  \omega)
}
\nonumber\\[1ex]&&\hspace{6ex}
      +\,t_{n/0}^q  \hat{\mathrm{E}}_{q\,{\rm in}}^{(n)}
      ({\bf k  },  \omega) +
      \hat{\mathrm{F}}_{q}^{(0)}({\bf k  },  \omega)
      \end{eqnarray}
and
\begin{eqnarray}
      \label{A2}
 \lefteqn{
     \hat{\mathrm{E}}_{q\,{\rm out}} ^{(n)} ({\bf k  }, \omega)
     = t^q_{0/n} \hat{\mathrm{E}}_{q\,{\rm in}} ^{(0)}({\bf k  },  \omega)
}
\nonumber\\[1ex]&&\hspace{6ex}
      +\,
      r_{n/0}^q  \hat{\mathrm{E}}_{q\,{\rm in}}^{(n)}
      ({\bf k  },  \omega) +
      \hat{\mathrm{F}}_{q}^{(n)}({\bf k  },  \omega),
      \end{eqnarray}
where
%\begin{widetext}
%\begin{eqnarray}
\begin{equation}
\begin{split}
	\label{A3}
%        \lefteqn{	
	&\hat{\mathrm{F}}_{q}^{(0)}({\bf k  },  \omega)  
%} \nonumber\\[1ex]&&
	= \sum _{j = 1 } ^{n-1}
        \left[ \phi _{q\, 0 +} ^{(j)}
        \hat{\mathrm{E}}_{q+}^{(j)}({\bf k  }, \omega) +
        \phi _{q\, 0 -} ^{(j)} \hat{\mathrm{E}}_{q-}^{(j)}
        ({\bf k  }, \omega) \right] \\
%}\nonumber\\[1ex]&&
	&= i \omega \mu _0 {\bf e}^{(0)}_{q-}({\bf k})
%%\nonumber\\[2ex]&& \; 	
        \cdot\sum _{j=1}^{n-1} \int _{[j]} \D z\,
        \mbb{g}^{(0j)}(0, z, {\bf k  }, \omega)\cdot
        \hat {\bf j}^{(j)} (z, {\bf k  }, \omega ),
%        \quad
%\nonumber\\&&
\end{split}
\end{equation}        
%\end{eqnarray}
%\begin{eqnarray}
\begin{equation}
\begin{split}
        \label{A4}
%        \lefteqn{
	&\hat{\mathrm{F}}_{q}^{(n)}({\bf k  },  \omega) 
%} \nonumber\\[1ex]&&	
	= \sum _{j = 1 } ^{n-1} \left[ \phi _{q\, n +} ^{(j)}
        \hat{\mathrm{E}}_{q+}^{(j)}({\bf k  }, \omega) +
        \phi _{q\, n -} ^{(j)} \hat{\mathrm{E}}_{q-}^{(j)}
        ({\bf k  }, \omega) \right]\\
%}\nonumber\\[1ex]&&
	&= i \omega \mu _0 \,{\bf e}^{(n)}_{q+}({\bf k})
%%\nonumber\\[2ex]&& \;       
	\cdot\sum _{j=1}^{n-1} \int _{[j]} \D z\,
        \mbb{g}^{(nj)}(0, z, {\bf k  }, \omega)\cdot
        \hat {\bf j}^{(j)} (z, {\bf k  }, \omega ).
%        \quad
%\nonumber\\&&        
%\end{eqnarray}
\end{split}
\end{equation}
%\end{widetext}        
Hence, the (relevant) commutators of the output amplitude operators that
refer to different sides of the plate can be given by 
\begin{widetext}
\begin{eqnarray}
      \label{A5}
\lefteqn{
      \left[
      \hat{\mathrm{E}}_{q\,{\rm out}} ^{(0)} ({\bf k  }, \omega),
      \hat{\mathrm{E}}_{q'\,{\rm out}} ^{(n)\dagger} ({\bf k  }', \omega')
      \right]
}
\nonumber\\[1ex]&&
       =\,r^q_{0/n} t_{0/n} ^{q\,*} 
       \left[
       \hat{\mathrm{E}}_{q\,{\rm in}} ^{(0)} ({\bf k  }, \omega),
       \hat{\mathrm{E}}_{q'\,{\rm in}} ^{(0)\dagger} ({\bf k  }', \omega')
       \right] 
       + t_{n/0} ^q r^{q\,*}_{n/0} 
       \left[
       \hat{\mathrm{E}}_{q\,{\rm in}} ^{(n)} ({\bf k  }, \omega),
       \hat{\mathrm{E}}_{q'\,{\rm in}} ^{(n)\dagger} ({\bf k  }', \omega')
       \right] 
       +
       \left[
       \hat{\mathrm{F}}_{q} ^{(0)} ({\bf k  }, \omega),
       \hat{\mathrm{F}}_{q'} ^{(n)\dagger} ({\bf k  }', \omega')
       \right] 
\nonumber\\[1ex]&&
       =\,\delta_{qq'}\, \delta(\omega-\omega')\,\delta({\bf k}-{\bf k}')\,
       \left( c_{q\,{\rm in}}^{(0)} r^q_{0/n} t_{0/n} ^{q\,*} 
	+
	c_{q\,{\rm in}}^{(n)}
	t_{n/0} ^q r^{q\,*}_{n/0} \right)
	+
	\left[
        \hat{\mathrm{F}}_{q} ^{(0)} ({\bf k  }, \omega),
        \hat{\mathrm{F}}_{q'} ^{(n)\dagger} ({\bf k  }', \omega')
        \right],
        \end{eqnarray}
%\end{widetext}
where Eqs.~(\ref{60}) -- (\ref{62}) have been used.
Making use of Eqs.~(\ref{17}) and (\ref{18})
and recalling the basic commutation relations (\ref{4})
and (\ref{5}), we derive
%\begin{widetext}
\begin{eqnarray}
      \label{A6}
\lefteqn{\left[
      \hat{\mathrm{F}}_{q} ^{(0)} ({\bf k  }, \omega)\,,\,
      \hat{\mathrm{F}}_{q'} ^{(n)\dagger} ({\bf k  }', \omega')
      \right]
}
\nonumber\\*[1ex]&&
	 =\,\delta_{qq'}\,\delta(\omega-\omega')\,
         \delta({\bf k}-{\bf  k}')\,
         \frac{4\pi \hbar}{\epsilon _0}\left(\frac {\omega}{c}\right) ^2
         \, e^{(0)}_{q-\, \mu} ({\bf k  })\,
         e^{(n)\, *}_{q+\, \mu '} ({\bf k  })
         \sum_{j=1}^{n-1}\,\int_{[j]} \D z\,
         g^{(0j)}_{\mu\nu}(0, z, {\bf k}, \omega )\,
         \frac {\omega ^2} {c^2} \varepsilon_{j}''\,
         g^{(nj)*}_{\mu'\nu}(0, z, {\bf k}, \omega )
\nonumber\\[1ex]&&
	 =\, \delta_{qq'}\,\delta(\omega-\omega')\,
         \delta({\bf k}-{\bf  k}')\,
         \frac{4\pi \hbar}{\epsilon _0}\left(\frac {\omega}{c}\right) ^2
         \, e^{(0)}_{q-\, \mu} ({\bf k  })\,  
         e^{(n)\, *}_{q+\, \mu '} ({\bf k  }) \,
	 \Biggl\lbrace \sum_{j=0}^{n}\,\int_{[j]} \D z\,
         g^{(0j)}_{\mu\nu}(0, z, {\bf k}, \omega )\,
         \frac {\omega ^2} {c^2} \varepsilon_{j}''\,
         g^{(nj)*}_{\mu'\nu}(0, z, {\bf k}, \omega ) 
	 \Biggr.
\nonumber\\[1ex]&&\qquad
         \Biggl.
	 - \int_{-\infty}^{0} \D z\,
         g^{(00)}_{\mu\nu}(0, z, {\bf k}, \omega )\,
         \frac {\omega ^2} {c^2} \varepsilon_{0}''\,
         g^{(n0)*}_{\mu'\nu}(0, z, {\bf k}, \omega )
	 - \int^{\infty}_{0} \D z\,
         g^{(0n)}_{\mu\nu}(0, z, {\bf k}, \omega )\,
         \frac {\omega ^2} {c^2} \varepsilon_{n}''\,
         g^{(nn)*}_{\mu'\nu}(0, z, {\bf k}, \omega )
         \Biggr\rbrace.
\end{eqnarray}
\end{widetext}
It is not difficult to calculate the last two integrals
in Eq.~(\ref{A6}), by using the explicit expression (\ref{20}) for
$\mbb{g}^{(jj')}(z, z', {\bf k  },  \omega )$.
To calculate the sum of integrals $\sum_{j=0}^{n}\,\int_{[j]} \D z\,\ldots$,
we employ the integral relation (\ref{11}) for the classical Green tensor
$\mbb{G} ({\bf r  }, {\bf r '}, \omega)$, rewritten in terms of 
$\mbb{g}^{(jj')}(z, z', {\bf k  }, \omega)$,
\begin{widetext}
\begin{eqnarray}
      \label{A7}
\lefteqn{
      \sum_{j''=0}^n\,\int_{[j'']} \D z''\,
      g^{(jj'')}_{\mu\nu}(z, z'', {\bf k}, \omega )\,
      \frac {\omega ^2} {c^2} \varepsilon_{j''}''\,
      g^{(j'j'')*}_{\mu'\nu}(z', z'', {\bf k}, \omega )
      =  \frac{1}{2i}
      g^{(jj')}_{\mu\mu'}(z, z', {\bf k}, \omega )
}
\nonumber\\[1ex]&&
      -\,\frac{1}{2i}
      g^{(j'j)*}_{\mu'\mu}(z', z, {\bf k}, \omega )
      + g^{(jj')}_{\mu\nu}(z, z', {\bf k}, \omega ) e_{z\,\nu}
      \,\frac{\varepsilon_{j'}''}{\varepsilon_{j'}^*}\,e_{z\, \mu'}
      + e_{z\,\mu} \,
      \frac{\varepsilon_{j}''}{\varepsilon_{j}}\,
      g^{(j'j)*}_{\mu'\nu}(z', z, {\bf k}, \omega ) e_{z\,\nu}
      \,.
      \end{eqnarray}
\end{widetext}
After lengthy, but straightforward calculations we then derive, on using
Eqs.~(\ref{61}) and (\ref{62}),
\begin{eqnarray}
      \label{A8}
\lefteqn{
   \left[  \hat{\mathrm{E}}_{s\,{\rm out}} ^{(0)} ({\bf k  },\omega),
   \hat{\mathrm{E}}_{p\,{\rm out}} ^{(n)\dagger} ({\bf k  }', \omega')
   \right]
}
\nonumber\\[1ex]&&   
   = \left[  \hat{\mathrm{E}}_{p\,{\rm out}} ^{(0)} ({\bf k  },\omega),
   \hat{\mathrm{E}}_{s\,{\rm out}} ^{(n)\dagger} ({\bf k  }', \omega')
   \right]=0,
\end{eqnarray} 
\begin{eqnarray}
\label{A9}
\lefteqn{
   \left[  \hat{\mathrm{E}}_{s\,{\rm out}} ^{(0)} ({\bf k  },\omega),
   \hat{\mathrm{E}}_{s\,{\rm out}} ^{(n)\dagger} ({\bf k  }', \omega')
   \right]
   = \delta(\omega-\omega')\,\delta({\bf k}-{\bf k}')
}\qquad
\nonumber\\[1ex]&&\times\;
   \frac{\pi \hbar}{\epsilon _0}\left(\frac {\omega}{c}\right)^2
   \left( \frac{i \beta_0 ''}{\abs{\beta _0 }^2} t^{s\,*}_{0/n}-
   \frac {i \beta _n''}{\abs{\beta _n }^2} t_{n/0}^{s}\right),
   \qquad
\end{eqnarray}
\begin{widetext}
\begin{eqnarray}
\label{A10}
\lefteqn{
   \left[  \hat{\mathrm{E}}_{p\,{\rm out}} ^{(0)} ({\bf k  },\omega),
   \hat{\mathrm{E}}_{p\,{\rm out}} ^{(n)\dagger} ({\bf k  }', \omega')
   \right]=
   \frac{\pi \hbar}{\epsilon _0}\left(\frac {\omega}{c}\right) ^2
   \left\{
   \frac {t_{n/0} ^p}{\beta _n \abs{k_n} ^2}
   \left(k^2\frac{k_n ^2}{k_n^{*\,2}} - \abs{\beta _n }^2 \right) 
   \right.
} 
\nonumber\\[1ex]&&
   \left.
   +\,\frac {t_{0/n} ^{p\, *}}{\beta _0 ^* \abs{k_0} ^2}
   \left(k^2\frac{k_0^{*\,2}}{k_0 ^2} - \abs{\beta _0 }^2 \right) 
   - \frac{\beta_n'}{\abs{\beta_n}^2}\,t^{p}_{n/0}
   \left[{\bf e}^{(n)}_{p-} ({\bf k  })\cdot{\bf e}^{(n)*}_{p-} ({\bf k  })\right]
   \left[{\bf e}^{(n)*}_{p-} ({\bf k})\cdot\,{\bf e}^{(n)*}_{p+} ({\bf k  })\right]
   \right.
\nonumber\\[1ex]&&
   -\, \frac{\beta_0'}{\abs{\beta_0}^2}\,t^{q*}_{0/n}\,
   \left[{\bf e}^{(0)*}_{q+} ({\bf k  })\cdot{\bf e}^{(0)}_{q+} ({\bf k  })\right]
   \left[{\bf e}^{(0)}_{q+} ({\bf k  })\cdot{\bf e}^{(0)}_{q-} ({\bf k  })\right]
   \Biggr\}\,
   \delta(\omega-\omega')\,\delta({\bf k}-{\bf k}')\,. 	
\end{eqnarray}
\end{widetext}
%%\begin{widetext}
%\begin{eqnarray}
%\label{A10}
%\lefteqn{
%   \left[  \hat{\mathrm{E}}_{p\,{\rm out}} ^{(0)} ({\bf k  },\omega),
%   \hat{\mathrm{E}}_{p\,{\rm out}} ^{(n)\dagger} ({\bf k  }', \omega')
%   \right]
%%} \nonumber\\[1ex]&&	
%	= \delta(\omega-\omega')\,\delta({\bf k}-{\bf k}')
%} \nonumber\\[1ex]&&	
%   \times
%   \frac{\pi \hbar}{\epsilon _0}\left(\frac {\omega}{c}\right) ^2
%   \Biggl\{
%   \frac {t_{n/0} ^p}{\beta _n \abs{k_n} ^2}
%   \left(k^2\frac{k_n ^2}{k_n^{*\,2}} - \abs{\beta _n }^2 \right) 
%   \Biggr. 
%\nonumber\\[1ex]&&
%   \left.
%\;   +\,\frac {t_{0/n} ^{p\, *}}{\beta _0 ^* \abs{k_0} ^2}
%   \left(k^2\frac{k_0^{*\,2}}{k_0 ^2} - \abs{\beta _0 }^2 \right)  \right.
%\nonumber\\[1ex]&&
%	\left.
%\;     -\, \frac{\beta_n'}{\abs{\beta_n}^2}\,t^{p}_{n/0}
%   \left[{\bf e}^{(n)}_{p-} ({\bf k  })\cdot{\bf e}^{(n)*}_{p-} ({\bf k  })\right]
%   \left[{\bf e}^{(n)*}_{p-} ({\bf k})\cdot\,{\bf e}^{(n)*}_{p+} ({\bf k  })\right]
%   \right.
%\nonumber\\[1ex]&&
%   \Biggl.
%\;      -\, \frac{\beta_0'}{\abs{\beta_0}^2}\,t^{q*}_{0/n}\,
%   \left[{\bf e}^{(0)*}_{q+} ({\bf k  })\cdot{\bf e}^{(0)}_{q+} ({\bf k  })\right]
%   \left[{\bf e}^{(0)}_{q+} ({\bf k  })\cdot{\bf e}^{(0)}_{q-} ({\bf k  })\right]
%   \Biggr\}\,.
%\nonumber\\ 	
%\end{eqnarray}
%%\end{widetext}

Next, let us consider the commutators of the output
amplitude operators that refer to the same sides of the plate.
Performing the same steps as before, we now arrive at 
\begin{eqnarray}
   \label{A11}
\lefteqn{
   \left[  \hat{\mathrm{E}}_{s\,{\rm out}} ^{(0)} ({\bf k  },\omega),
   \hat{\mathrm{E}}_{p\,{\rm out}} ^{(0)\dagger} ({\bf k  }', \omega')
   \right]
}
\nonumber\\[1ex]&&
   = \left[  \hat{\mathrm{E}}_{p\,{\rm out}} ^{(0)} ({\bf k  },\omega),
   \hat{\mathrm{E}}_{s\,{\rm out}} ^{(0)\dagger} ({\bf k  }', \omega')
   \right] = 0,
\end{eqnarray}
\begin{eqnarray}
   \label{A12}
\lefteqn{
   \left[  \hat{\mathrm{E}}_{s\,{\rm out}} ^{(0)} ({\bf k  },\omega),
   \hat{\mathrm{E}}_{s\,{\rm out}} ^{(0)\dagger} ({\bf k  }', \omega')
   \right]
   = \delta(\omega-\omega')\,\delta({\bf k}-{\bf k}')
}\qquad
\nonumber\\[1ex]&&\times\;
   \frac{\pi \hbar}{\epsilon _0}\left(\frac {\omega}{c}\right) ^2
   \left( \frac{\beta_0 '}{\abs{\beta _0 }^2} +
   \frac {2 \beta _0''}{\abs{\beta _0 }^2} r_{0/n}^{s\,\prime \prime}\right),
   \qquad
\end{eqnarray}
\begin{widetext}
\begin{eqnarray}
   \label{A13}
\lefteqn{
   \left[  \hat{\mathrm{E}}_{p\,{\rm out}} ^{(0)} ({\bf k  },\omega),
   \hat{\mathrm{E}}_{p\,{\rm out}} ^{(0)\dagger} ({\bf k  }', \omega')
   \right]= \frac{\pi \hbar}{\epsilon _0}\left(\frac{\omega}{c}\right) ^2
   \left\{
   \frac {r_{0/n} ^p}{\beta _0 \abs{k_0} ^2}
   \left(k^2\frac{k_0 ^2}{k_0^{*\,2}} - \abs{\beta _0 }^2 \right)
   \right.
}
\nonumber\\[1ex]&& 
   \left.
   +\,\frac {r_{0/n} ^{p\, *}}{\beta _0 ^* \abs{k_0} ^2}
   \left(k^2\frac{k_0^{*\,2}}{k_0 ^2} - \abs{\beta _0 }^2 \right)
   - \frac {k ^2}{ \abs{k_0} ^2}\left(\frac{\beta _0}{k_0^{*\,2}}+
   \frac{\beta _0 ^*}{k_0^{2}}  \right)
   + \frac {2 \beta _0 '}{\abs{k_0} ^2}\left( \frac {k ^4}
   {\abs{\beta_0} ^2 \abs{k_0} ^2 } + 1 \right) 
   \right.
\nonumber\\[1ex]&& 
   +\,\frac{\beta_0 '}{\abs{\beta _0 }^2}
   \left[ {\bf e}^{(0)}_{p+} ({\bf k  })
   \cdot{\bf e}^{(0)*}_{p+} ({\bf k  })\right]
   \left[ \left| r_{0/n}^q\right|^2
   - \left|{\bf e}^{(0)}_{p-} ({\bf k })\cdot
   {\bf e}^{(0)}_{p+} ({\bf k  }) + r_{0/n} ^p\right| ^2 \right] 
   \Biggr\}\,
   \delta(\omega-\omega')\,\delta({\bf k}-{\bf k}').
\end{eqnarray}
\end{widetext}
%%\begin{widetext}
%\begin{eqnarray}
%   \label{A13}
%\lefteqn{
%   \left[  \hat{\mathrm{E}}_{p\,{\rm out}} ^{(0)} ({\bf k  },\omega),
%   \hat{\mathrm{E}}_{p\,{\rm out}} ^{(0)\dagger} ({\bf k  }', \omega')
%   \right]=  \delta(\omega-\omega')\,\delta({\bf k}-{\bf k}') 
%   } \nonumber\\[1ex]&&	
%   \times
%	\frac{\pi \hbar}{\epsilon _0}\left(\frac{\omega}{c}\right) ^2
%   \Biggl\{
%   \frac {r_{0/n} ^p}{\beta _0 \abs{k_0} ^2}
%   \left(k^2\frac{k_0 ^2}{k_0^{*\,2}} - \abs{\beta _0 }^2 \right)
% \Biggr.
% \nonumber\\[1ex]&& 
%   \left.
% \;    \, +\,\frac {r_{0/n} ^{p\, *}}{\beta _0 ^* \abs{k_0} ^2}
%   \left(k^2\frac{k_0^{*\,2}}{k_0 ^2} - \abs{\beta _0 }^2 \right) \right.
% \nonumber\\[1ex]&&
%	\left.
%\;  - \frac {k ^2}{ \abs{k_0} ^2}\left(\frac{\beta _0}{k_0^{*\,2}}+
%   \frac{\beta _0 ^*}{k_0^{2}}  \right)
%   + \frac {2 \beta _0 '}{\abs{k_0} ^2}\left( \frac {k ^4}
%   {\abs{\beta_0} ^2} + 1 \right) 
%   \right.
%\nonumber\\[1ex]&& 
%   \;   +\,  
%   \frac{\beta_0 '}{\abs{\beta _0 }^2}
%   \left[ {\bf e}^{(0)}_{p+} ({\bf k  })
%   \cdot{\bf e}^{(0)*}_{p+} ({\bf k  })\right]
%\nonumber\\[1ex]&& \Biggl. \times
%   \left[ \left| r_{0/n}^q\right|^2
%   - \left|\left[ {\bf e}^{(0)}_{p-} ({\bf k })\cdot
%   {\bf e}^{(0)}_{p+} ({\bf k  })\right] + r_{0/n} ^p\right| ^2 \right] 
%   \Biggr\}\,.
%%\nonumber\\ 
%\end{eqnarray}
%%\end{widetext}
The commutators $[\hat{\mathrm{E}}_{q\,{\rm out}} ^{(n)} ({\bf k  },\omega),
\hat{\mathrm{E}}_{q'\,{\rm out}} ^{(n)\dagger} ({\bf k  }', \omega')]$
are obtained from Eqs.~(\ref{A11}) -- (\ref{A13}), by making the
replacements
$\beta_0$ $\!\to$ $\!\beta_n$,
$k_0$ $\!\to$ $\!k_n$,
${\bf e}_{q\pm}^{(0)}$ $\!\to$ $\!{\bf e}_{q\mp}^{(n)}$,
and
$r_{0/n}^q$ $\!\to$ $\!r_{n/0}^q$.

Finally, it can easily be proved that the intraplate amplitude
operators (\ref{38}) satisfy the commutation relations
(\mbox{$j$ $\!=$ $\!1,\ldots,n$ $\!-$ $\!1$};
\mbox{$\lambda$ $\!=$ $\!\pm$})
\begin{eqnarray}
\label{A14}
\lefteqn{
   \left[\hat{E}_{q\lambda}^{(j)}({\bf k},\omega),
   \hat{E}_{q\lambda'}^{(j)\dagger}({\bf k}',\omega')\right]
}
\nonumber\\[1ex]&&
   = c_{q\lambda\lambda'}^{(j)}({\bf k},\omega)\,
   \delta_{qq'}\,\delta(\omega-\omega')\,\delta({\bf k}-{\bf k}'),
\end{eqnarray} 
where
\begin{eqnarray}
      \label{A15}
\lefteqn{
      c_{q\,{\rm \pm \pm  }}^{(j)} = \pm \frac{\pi \hbar}{\epsilon_0}
      \left(\frac{\omega}{c}\right)^2\frac{\beta_{j}'}{|\beta_{j}|^2}
}
\nonumber\\[1ex] &&\times\;
      \left( e^{\pm 2\beta _j '' d_j} - 1 \right)
      {\bf e}^{(j)}_{q \pm} ({\bf k  })
      \cdot{\bf e}^{(j)*}_{q\pm} ({\bf k}),
\\[1ex]
      \label{A16}
\lefteqn{
     c_{q\,{\rm \pm \mp }}^{(j)} = \pm i\frac{\pi \hbar}{\epsilon_0}
     \left(\frac{\omega}{c}\right)^2
     \frac{\beta_{j}''}{|\beta_{j}|^2}
}
\nonumber\\[1ex] &&\times\;
   \left( e^{\mp 2i \beta _j ' d_j} - 1 \right)
   {\bf e}^{(j)}_{q \pm} ({\bf k  })
   \cdot{\bf e}^{(j)*}_{q\mp} ({\bf   k}).
\end{eqnarray} 
Using Eqs.~(\ref{A14}) -- (\ref{A16}), we find that the operators 
\begin{eqnarray}
\lefteqn{
   \hat{a}_{q\pm}^{(j)} ({\bf k  }, \omega)
   = \frac {1} {\xi^{(j)}_{q \pm}({\bf k  }, \omega)}    
}
\nonumber\\[1ex]&&\times\;  
   \left[ e^{i \beta_{j} d_{j}}
   \hat{\mathrm{E}}_{q+}^{(j)} ({\bf k  }, \omega)  \pm
   \hat{\mathrm{E}}_{q-}^{(j)} ({\bf k  }, \omega) \right], 
   \label{A17}
\end{eqnarray}
where 
\begin{eqnarray}
   \label{A18}
\lefteqn{
   \xi^{(j)}_{q \pm}({\bf k  }, \omega) = 
   \frac  {2  \omega  } {c \beta _j }\sqrt{\frac {\pi \hbar } {\epsilon _0}}
   \,e^{ - \beta _{j}'' d_{j} /2}\,
}
\nonumber\\[1ex]&&\,\times\;
   \left\{
   \beta _j ' \sinh(\beta_j ''d_j)\,
   {\bf{e}}^{(j)}_{q+}({\bf k})
   \cdot {\bf{e}}^{(j)\,*}_{q+}({\bf k})
   \right.
\nonumber\\[1ex]&&\hspace{2ex} 
   \left.
   \pm\, \beta _j ''
   \sin(\beta_j 'd_j)\,
   {\bf{e}}^{(j)}_{q+}({\bf k})
   \cdot {\bf{e}}^{(j)\,*}_{q-}({\bf k})
   \right\} ^{1/2}\!\qquad
\end{eqnarray}
satisfy bosonic commutation relations. From Eq.~(\ref{A17}) it
then follows that
\begin{equation}
\label{A19}
\hat{E}_{q\lambda}^{(j)}({\bf k},\omega)
= \sum_{\lambda'=\pm} \tau_{q\lambda\lambda'}^{(j)}({\bf k},\omega)
\hat{a}_{q\lambda}^{(j)}({\bf k},\omega), 
\end{equation}
where
\begin{eqnarray}
\label{A20}
&&\tau_{q + \pm}^{(j)} =
{\textstyle\frac {1} {2}}
\xi^{(j)}_{q \pm}({\bf k  }, \omega) \,e^{-i \beta _j d_j},  
\\[1ex]
\label{A21}
&&
\tau_{q - \pm}^{(j)} = \pm {\textstyle\frac {1} {2}}
\xi^{(j)}_{q \pm}({\bf k  }, \omega).
\end{eqnarray}

In the limiting case \mbox{$\varepsilon_{0,n}$ $\!\to$ $\!0$}
(when the plate is surrounded by vacuum)
one has to distinguish between \mbox{$\omega/c$ $\!>$ $\!k$}
(propagating-field components) and \mbox{$\omega/c$ $\!\leq$ $\!k$}
(evanescent-field components). In the first case
\mbox{($\omega/c$ $\!>$ $\!k$)} we have
\mbox{$\beta_0'$ $\!=$ $\!\beta_0$ $\!=$ $\!\beta_n$ $\!=$
$\!\beta_n'$ $\!>0$}, so that Eqs.~(\ref{A8}) -- (\ref{A10})
reduce to 
\begin{equation}
      \label{A22}
      \left[
      \hat{\mathrm{E}}_{q\,{\rm out}} ^{(0)} ({\bf k  }, \omega),
      \hat{\mathrm{E}}_{q'\,{\rm out}} ^{(n)\dagger} ({\bf k  }', \omega')
      \right]
      = 0,
\end{equation}
and Eqs.~(\ref{A11}) -- (\ref{A13}) simplify to
\begin{eqnarray}
      \label{A23}
\lefteqn{
      \left[
      \hat{\mathrm{E}}_{q\,{\rm out}} ^{(0)} ({\bf k  }, \omega),
      \hat{\mathrm{E}}_{q'\,{\rm out}} ^{(0)\dagger} ({\bf k  }', \omega')
      \right]
}\nonumber\\[1ex]&&
      =\,\frac{\pi \hbar}{\epsilon_0}\left(\frac{\omega}{c}\right)^2
      \frac{1}{\beta_0}\,\delta_{qq'}\,\delta(\omega-\omega')\,
      \delta({\bf k}-{\bf k}').\qquad
      \end{eqnarray}  
\begin{eqnarray}
      \label{A24}
\lefteqn{
      \left[
      \hat{\mathrm{E}}_{q\,{\rm out}} ^{(n)} ({\bf k  }, \omega),
      \hat{\mathrm{E}}_{q'\,{\rm out}} ^{(n)\dagger} ({\bf k  }', \omega')
      \right]
}
\nonumber\\[1ex]&&
      =\,\frac{\pi \hbar}{\epsilon_0}\left(\frac{\omega}{c}\right)^2
      \frac{1}{\beta_n}\,\delta_{qq'}\,
      \delta(\omega-\omega')\,\delta({\bf k}-{\bf  k}').\qquad
      \end{eqnarray}
In the second case
\mbox{($\omega/c$ $\!\leq$ $\!k$)} we have
\mbox{$\beta_0'$ $\!=$ $\!\beta_n'$ $\!=0$}. Equations
(\ref{A8}) -- (\ref{A10}) then lead to  
\begin{eqnarray}
      \label{A25}
\lefteqn{
      \left[ \hat{\mathrm{E}}_{q\,{\rm out}} ^{(0)} ({\bf k  }, \omega),
      \hat{\mathrm{E}}_{q'\,{\rm out}} ^{(n)\dagger} ({\bf k  }',
      \omega') \right]
}
\nonumber\\[1ex]&&
      =\,\frac{\pi \hbar}{\epsilon_0}
      \left(\frac{\omega}{c}\right)^2
      \frac{2 t^{q\, \prime \prime}_{0/n}}{\abs{\beta_0}}
      \,\delta_{qq'}\, \delta(\omega-\omega')\,\delta({\bf k}-{\bf  k}'),
      \qquad
\end{eqnarray}
and from Eqs.~(\ref{A11}) -- (\ref{A13}) we find
\begin{eqnarray}
      \label{A26}
\lefteqn{
      \left[
      \hat{\mathrm{E}}_{q\,{\rm out}} ^{(0)} ({\bf k  }, \omega),
      \hat{\mathrm{E}}_{q'\,{\rm out}} ^{(0)\dagger} ({\bf k  }', \omega')
      \right]
}
\nonumber\\[1ex]&&
      =\,\frac{\pi \hbar}{\epsilon_0}
      \left(\frac{\omega}{c}\right)^2
      \frac{2 r_{0/n}^{q\,\prime \prime}}{\abs{\beta_0}}\,
      \delta_{qq'}\,\delta(\omega-\omega')\,\delta({\bf k}-{\bf  k}'),
      \qquad
\end{eqnarray}
and similarly
\begin{eqnarray}
      \label{A27}
\lefteqn{
      \left[
      \hat{\mathrm{E}}_{q\,{\rm out}} ^{(n)} ({\bf k  }, \omega),
      \hat{\mathrm{E}}_{q'\,{\rm out}} ^{(n)\dagger} ({\bf k  }', \omega')
      \right]
}
\nonumber\\[1ex]&&
      =\,\frac{\pi \hbar}{\epsilon_0}
      \left(\frac{\omega}{c}\right)^2
      \frac{2 r_{n/0}^{q\,\prime \prime}}{\abs{\beta_n}}\,
      \delta_{qq'}\,\delta(\omega-\omega')\,\delta({\bf k}-{\bf  k}').
      \qquad
\end{eqnarray}

Note that the commutation
relations (42) in Ref.~\cite{Savasta19} for the
bosonic output operators are only valid for propagating
waves, not for evanescent ones. It should also
be noted that the free-space light modes
${\bf U}_{p,{\bf K}}(z,\omega)$ used in Ref.~\cite{DiStefano01}
are introduced by explicitly requiring that
\mbox{$\omega/c$ $\!>$ $\!|{\bf K}|$},
where ${\bf K}$ corresponds to ${\bf k}$ in the present paper.
Hence, these modes solely represent propagating waves. 

%%%%%%%%%%%%%%%%%%%%%%%%%%%%%%%%%%%%%%%%%%%%%%%%%%%%%%%%%%%%%%%%%%%%%%%%%%%
%%%%%%%%%%%%%%%%%%%%%%%%%%%%%%%%%%%%%%%%%%%%%%%%%%%%%%%%%%%%%%%%%%%%%%%%%%%

%\end{thebibliography}
\end{document}